\documentclass[12pt,a4paper,twoside,groupcitations]{article}
\usepackage[T1]{fontenc}
\usepackage[ansinew]{inputenc}
\usepackage[english]{babel}
\usepackage{amsfonts}
\usepackage{amsmath}
\usepackage{array}
\usepackage{amsthm}
\usepackage{amssymb}
\usepackage{graphicx}
\usepackage{subfigure}
\usepackage{braket}
\usepackage{eucal}
\usepackage{verbatim}
\usepackage[table]{xcolor}
\usepackage{caption}
\usepackage{cite}
\usepackage{textcomp}
\usepackage{mathtools}
\usepackage[multiple]{footmisc}
\usepackage{multicol}
\usepackage{tikz}
\usetikzlibrary{positioning,arrows}
\usetikzlibrary{decorations.pathmorphing}
\usetikzlibrary{decorations.markings}
\usetikzlibrary{calc,decorations.markings}
\usetikzlibrary{arrows,shapes}
\usetikzlibrary{matrix,arrows}
\usepackage{pgfplots}
\usepackage{xparse}
\definecolor{jade}{HTML}{00A86B}
\newcommand{\be}{\begin{eqnarray}}
\newcommand{\ee}{\end{eqnarray}}
\newcommand{\pro}[2]{\mbox{$\langle\, #1 \mid #2\,\rangle$}}

\renewcommand{\d}{\mbox{${\rm d}$}} 

\newcommand{\gn}{G_{\rm N}}

\newcommand{\beq}{\begin{equation}}
\newcommand{\eeq}{\end{equation}}
\DeclareMathOperator{\Tr}{Tr}

\title{\bf Covariant singularities in Quantum Field Theory and Quantum Gravity}
\author{Roberto~Casadio$^{ab}$\thanks{E-mail: casadio@bo.infn.it},
$\ $
Alexander~Kamenshchik$^{abc}$\thanks{E-mail: kamenshchik@bo.infn.it},
$\ $
and
Iber\^e Kuntz$^{ab}$\thanks{E-mail: kuntz@bo.infn.it}
\\
\\
$^a${\em Dipartimento di Fisica e Astronomia, Universit\`a di Bologna}
\\
{\em via Irnerio~46, 40126 Bologna, Italy}
\\
\\
$^b${\em I.N.F.N., Sezione di Bologna, I.S.~FLAG}
\\
{\em viale B.~Pichat~6/2, 40127 Bologna, Italy}
\\
\\
$^c${\em L.D.~Landau Institute for Theoretical Physics}
\\
{\em of the Russian Academy of Sciences} 
\\
{\em 119334 Moscow, Russia}
}
\begin{document}
\maketitle
\begin{abstract}
It is rather well-known that spacetime singularities are
not covariant under field redefinitions.
A manifestly covariant approach to singularities in classical gravity was proposed
in~\cite{Casadio:2020zmn}.
In this paper, we start to extend this analysis to the quantum realm.
We identify two types of covariant singularities in field space corresponding to geodesic
incompleteness and ill-defined path integrals (hereby dubbed functional singularities).
We argue that the former might not be harmful after all, whilst the latter makes all observables
undefined.
We show that the path-integral measure is regular in any four-dimensional theory
of gravity without matter or in any theory in which gravity is either absent or treated semi-classically.
This might suggest the absence of functional singularities in these cases,
however it can only be confirmed with a thorough analysis, case by case, of the path integral.
We provide a topological and model-independent classification of functional singularities
using homotopy groups and we discuss examples of theories with and without such singularities.
\end{abstract}
\section{Introduction}
\setcounter{equation}{0}
\label{Sintro}
The complexity of Nature and its infinitude of details give rise to an uncountable number
of possibilities to describe it using physical models.
From a practical point of view, narrowing down these possibilities amounts to the almost
artistic job of identifying relevant hypotheses and then testing them against observations.
Along the history of physics, this process has resulted in very important fundamental
principles, such as coordinate-independence and gauge invariance.
With the advent of modern geometry, the former has practically become a no-brainer
and has achieved such a level of confidence that it is usually taken for granted.
Gauge invariance under Lie symmetries has similarly gained a lot of trust after
culminating in the Standard Model of particle
physics~\cite{Yang:1954ek,Glashow:1961tr,Weinberg:1967tq,Salam:1968rm}.
Geometrically, gauge invariance does not seem much different from coordinate-independence,
as it reflects the ambiguity in choosing coordinates (or more precisely local trivialization)
of the vector bundle.
Both notions definitely share the same idea that physics should not depend on 
the artificial choices we make to describe it.
\par
There is a third instance of this seemingly general idea of coordinate-independence,
namely the ambiguity in the choice of fields.
Albeit not as popular as the other two, field redefinitions play prominent roles in both
classical and quantum field theories.
Their application in gravity has led to the discovery of both Starobinsky's model of
inflation~\cite{Starobinsky:1979ty,Starobinsky:1980te}, whose inflaton is hidden in the
square of the Ricci curvature, and Higgs inflation~\cite{Bezrukov:2007ep,Barvinsky:2008ia},
which results from a field redefinition of the non-minimal coupling of the Higgs
boson to gravity.
Field redefinitions have also been employed to show the equivalence between
particle dark matter and some models of modified gravity in which additional degrees of freedom are present~\cite{Calmet:2017voc}.
In quantum field theory, field redefinitions have been used for simplifying the
renormalization procedure by taming UV
divergences~\cite{Solodukhin:2015ypa,Kazakov:1987ej,Mohammedi:2013oca,
Slovick:2013rya,Apfeldorf:1994av} and for simplifying the action of effective field
theories~\cite{Arzt:1993gz,Georgi:1991ch,Criado:2018sdb,Passarino:2016saj}.
This is in fact justified by Borchers' theorem~\cite{borchers,Parker:2009uva},
which states that $S$-matrix elements are invariant under field redefinitions
interpolating between fixed asymptotic states.
Nonetheless, there is no guarantee that more general field redefinitions
would leave the $S$-matrix invariant, let alone other quantities in the theory,
such as the effective action.
This problem is actually intimately related to the gauge-dependence of the
standard effective action, which motivated the introduction of a covariant effective
action under general field redefinitions by Vilkovisky and
DeWitt~\cite{Vilkovisky:1984st,DeWitt:1988dq,Ellicott:1987ir}.
More recently, field-covariant formulations of quantum field theory have regained
some attention~\cite{Anselmi:2012aq,Anselmi:2012qy,Finn:2019aip,Finn:2020nvn}.
\par
In spite of the usefulness of field redefinitions, their interpretation in gravity has not
reached a consensus yet.
There is in fact a long-standing debate concerning whether the Jordan frame or the Einstein
frame should be regarded as
physical~\cite{Kamenshchik:2014waa,Capozziello:2010sc,Capozziello:1996xg,
Postma:2014vaa,Faraoni:1999hp,Sk.:2016tvl,Pandey:2016unk}.
While some argue in favour of the former and some in favour of the latter, many others
hold the position that they are actually equivalent.
Ideally, this type of discussion should be framed in terms of physical observables.
Nevertheless, what comprises the list of observables in gravity is a rather subtle and
non-trivial subject {\em per se}.
The best way to phrase the equivalence (or non-equivalence) between different frames
is then through the action.
It turns out that the classical action $S=S[\phi]$, which describes the dynamics of a set
of classical fields $\phi$,
is a scalar under field redefinitions.
In fact, for any one-to-one transformation
\beq
\phi
\to
\tilde\phi
=f(\phi)
\ ,
\label{eq:1-1}
\eeq
the classical action transforms as
\beq
\tilde S[\tilde \phi] = S[\phi]
\ .
\eeq
Furthermore, the equations of motion transform covariantly,
\beq
\frac{\delta \tilde S[\tilde\phi]}{\delta\tilde\phi}
=
\frac{\delta\phi}{\delta\tilde\phi} \frac{\delta S[\phi]}{\delta\phi}
\ ,
\eeq
thus solutions in the frame $\phi$ are taken into solutions in the frame $\tilde\phi$
and vice-versa.~\footnote{We remark that this does not prevent the existence of
extremals of $S[\phi]$ that cannot be mapped into extremals of $\tilde S[\tilde\phi]$ 
when the mapping~\eqref{eq:1-1} is not regular.
This also calls for attention to boundary contributions in the definition of the action,
like the famous Gibbons-Hawking-Brown-York term~\cite{reggeT}.}
In this sense, $\phi$ and $\tilde\phi$ are just coordinates in the field space.
We have no means to tell them apart, regardless of whether the set of fields $\phi$ includes the
spacetime metric.
At the formal level, one could simply require that classical observables be scalars under field
redefinitions.
This demands, of course, great care when interpreting objects in a theory.
Some equations and quantities might very well look completely different in different
field-space coordinates.
Fundamental concepts might change or even lose their meaning.
Must we not forget that, in the end of the day, the fate of a theory is dictated by its
experimentally observable predictions, everything else is just instrumental 
to obtaining them.
\par
At the quantum level, on the other hand, the above considerations might change
quite considerably.
Notwithstanding Borchers' theorem, scattering amplitudes are not invariant under
general field redefinitions, but only those that keep the asymptotic states fixed.
This would suggest a preferred set of coordinate systems in field space.
Furthermore, $S$-matrix elements are not the only observables in a quantum field theory.
Correlation functions, that is in-in amplitudes, calculated in the Schwinger-Keldysh
formalism are the actual observables in cosmology and astrophysics.
The question of coordinate-independence in field space thus becomes of fundamental
importance for studying quantum effects in these contexts.
Contrary to the classical action, the effective action, which encodes all the information
of a quantum system, explicitly turns out not to be a scalar under field redefinitions.
This clearly opens up two possibilities as either i) physics depends on field-space coordinates
at the quantum level or ii) the effective action formalism is not complete.
\par
In this paper, we shall take the position that physics at the fundamental level
should not depend on the way fields are parameterised.
There are at least two important reasons for this. For one, as explained above,
the classical action does not depend on the field parameterization, thus it is reasonable
to keep this property in the quantum theory as well.
Secondly, and more importantly, a dependence on the parameterisation is suggestive
of an anthropic viewpoint, in which one attaches a special meaning to one's choice
(in this case, the choice of coordinates in field space), whereas no choice appears
physically favoured \textit{a priori}.
In fact, the results of any experiment are numerical readings that do not imply any particular
parameterisation of the field variables.
Such a special choice would thus be rather artificial.
\par
From this point of view, the effective action must be amended in order to be invariant
under field redefinitions.
Vilkovisky was the first to study the dependence of the effective action on a particular
choice of field-space coordinates~\cite{Vilkovisky:1984st}.
He introduced a connection in field space which is able to cancel the sources of
non-covariance and yet reduce the formalism to the usual effective action for a
vanishing connection.
Vilkovisky's formalism was latter generalised to curved field-spaces by
DeWitt~\cite{DeWitt:1988dq} and the resulting formalism goes by the name of
the Vilkovisky-DeWitt effective action.
Such an effective action is invariant under arbitrary field redefinitions, which guarantees
the invariance of all physical observables at the quantum level, including scattering
amplitudes and correlation functions.
\par
The purpose of this paper is to scrutinise singularities under field redefinitions
in quantum field theories and quantum gravity.
This complements our classical analysis of singularities under field redefinitions
in Ref.~\cite{Casadio:2020zmn}, where we have shown that the field-space
Kretschmann scalar for pure gravity is everywhere finite.
In particular, we shall show the possibility of two distinct types of singularities
in field space: geodesics incompleteness (named covariant singularities
in~\cite{Casadio:2020zmn}) and points in field space where the path integral is
not well-defined (which shall be called functional singularities).
The former leads to no serious issues as points in field space that are unreachable
by geodesics can be interpreted as non-existent configurations, whilst the latter
would make all physical observables undefined.
We shall also propose a practical way of identifying the existence of the latter,
more problematic kind of singularity, by explicitly computing winding numbers
in field space.
\par
The paper is organised as follows:
in Section~\ref{Sea}, we review the standard formalism of the effective action;
we introduce geometrical concepts of field space in Section~\ref{Sgeo} and
use them to construct the invariant effective action in Section~\ref{Saction},
following the works by Vilkovisky and DeWitt;
Section~\ref{Ssingularity} is devoted to the study of functional singularities,
their relation to spacetime singularities and strategies to overcome their
corresponding issues;
in Section~\ref{Stop}, we use homotopy groups to classify singularities
in field space and provide a few examples;
conclusions are drawn in Section~\ref{Sconc}.
\par
To avoid cluttered equations, we shall adopt DeWitt's condensed notation
(for more details, see Ref.~\cite{Parker:2009uva}):
as usual, mid-alphabet Greek letters (e.g.~$\mu,\nu,\rho, \ldots$)
shall denote spacetime indices;
lowercase mid-alphabet Latin indices (e.g.~$i, j, k, \ldots$) collectively represent
both discrete indices (denoted by the corresponding capital Latin letters $I, J, K, \ldots$),
and the continuum spacetime coordinates $x\equiv x^\mu$.~\footnote{Indices will be omitted
overall when no confusion can arise.}
This association can be schematically represented as $i = (I, x)$, so that $\phi^i = \phi^I(x)$
are the coordinates of a field configuration.
Repeated mid-alphabet lowercase indices results in summations over
all the discrete indices and integration over the spacetime $\Omega$
of dimension ${\rm dim}(\Omega)=n$.
We shall usually denote points $P\in\Omega$ with their coordinates.
For example, we will write $x\in\Omega$ as a shorthand notation for $x(P)\in U(\Omega)$, where
$U\in\mathbb R^n$ is the domain of the chart of $\Omega$ including the point $P$
of coordinates $x$.
Lowercase Latin indices of the beginning of the alphabet (e.g.~$a, b, c, \ldots$)
shall be reserved to internal indices in gauge theories and indices corresponding
to the beginning of the Greek alphabet (e.g.~$\alpha, \beta, \gamma, \ldots$)
will denote spinor indices.
\section{Quantum amplitudes and the effective action}
\setcounter{equation}{0}
\label{Sea}
Functional methods~\cite{DeWitt:1962ud,DeWitt:1965jb}, like the Schwinger action
principle and the Feynman path integral, concentrate on the transition amplitude
between two quantum states.
More specifically, the spacetime $\Omega$ is assumed to be globally hyperbolic and admits
foliations in (spatial) Cauchy hypersurfaces $\Sigma_t$, which we label by a time
coordinate $t$ for simplicity.
We shall also assume that all the fields $\phi$ are spacetime tensors and their action
$S$ is a scalar functional of them.
One then considers the transition amplitude between a certain quantum state
$\ket{\zeta_{\rm in},t_{\rm in}}$ defined on the initial hypersurface $\Sigma_{t_{\rm in}}$
and a given final state $\ket{\zeta_{\rm out},t_{\rm out}}$ on a different
hypersurface $\Sigma_{t_{\rm out}}$,~\footnote{One might be interested in the evolution
forward in time, so that $t_{\rm out}>t_{\rm in}$, or backward in time, with $t_{\rm out}<t_{\rm in}$.}
\be
\pro{\zeta_{\rm out},t_{\rm out}}{\zeta_{\rm in},t_{\rm in}}
=
\bra{\zeta_{\rm in},t_{\rm in}}
U^{-1}(t_{\rm out},t_{\rm in})
\ket{\zeta_{\rm in},t_{\rm in}}
\ ,
\label{ta}
\ee
where the unitary operator evolving states along the foliation is given by
\be
U({t_2,t_1})
=
\exp\left(-\frac{i}{\hbar}\,S[\phi;t_1,t_2]\right)
\ ,
\ee
with the action $S$ computed on the spacetime volume $\bar\Omega\subseteq\Omega$
between the hypersurfaces $\Sigma_{t_1}$ and $\Sigma_{t_2}$.
Since the transition amplitude~\eqref{ta} only depends on the initial and final states,
any variation $\delta\phi$ of the fields which vanishes on the initial and final
hypersurfaces must leave it unchanged.
This yields the operator equations of motion for the fields $\phi$ from the Schwinger
variational principle $\delta S=0$. 
Of course, the field equations for the operators $\phi$ do not depend on the states
$\ket{\zeta_{\rm in},t_{\rm in}}$ and $\ket{\zeta_{\rm out},t_{\rm out}}$, but one is
eventually interested in computing the transition amplitudes themselves or the
expectation values of quantum observables for the chosen quantum states.
\par
In Feynman approach, one computes how field excitations evolve between
the initial and final quantum states by first coupling each of the fields $\phi^i$
to an external current $J_i$, that is
\be
S_J[\phi]
=
S[\phi]
+ J_i\,\phi^i
\ .
\ee 
The insertion of these currents results in the path integral
\be
\pro{\zeta_{\rm out},t_{\rm out}}{\zeta_{\rm in},t_{\rm in}}[J]
&\!=\!&
\int
\left[\prod_i\d\phi^i\right]
\exp\left\{\frac{i}{\hbar}\,S_J[\phi]\right\}
\nonumber
\\
&\!\coloneqq\!&
\exp\left\{\frac{i}{\hbar}\,W[J]\right\}
\ ,
\label{Fp}
\ee
where all field configurations compatible with the boundary conditions at $\Sigma_{t_{\rm in}}$
and $\Sigma_{t_{\rm out}}$ are summed over.
Functional derivatives of $W[J]$ with respect to the currents $J_i$ (evaluated at $J_i=0$) then
yield correlation functions and quantum corrected equations of motion for the (expectation values of)
the fields $\phi^i$. 
In the functional integral~\eqref{Fp}, one can change how the fields $\phi$ are represented as long as
the redefinition is not singular and can be inverted. 
Hence, the very starting point of functional quantisation shows how deeply entangled are the geometry
of spacetime (represented by the foliation $\Sigma_t$ and the expectation values of the fields $\phi$)
and the geometry of field space in which the $\phi^i$ act as coordinates.
It should be noted that the path integral measure in Eq.~\eqref{Fp} is not invariant under
general field redefinitions. This is not a problem in perturbation theory because the Jacobian of the redefinition
is one in dimensional regularization.
However, the general definition of the functional measure is a quite intricate problem as different formalisms
yield different results~\cite{Leutwyler:1964wn,Fradkin:1973wke,Faddeev:1973zb,Fujikawa:1983im}.
For our purposes, an invariant measure will be required in order to make correlation functions covariant
under field redefinitions.
We shall study this issue in more depth in the next sections.
\par
In this construction, the role of the initial and final quantum states is crucial and should not be
overlooked.
For instance, high energy physicists are usually interested in scattering processes occurring in 
a (supposedly) well defined vacuum
$\ket{\zeta_{\rm in},t_{\rm in}\to-\infty}=\ket{\zeta_{\rm out},t_{\rm out}\to+\infty} \equiv\ket{0}$.
Moreover, the question one is usually trying to answer is the probability that incoming particles
of given momenta result in certain final states that can be experimentally detected.
Each scattering process is therefore characterised by certain types and numbers of incoming and
outgoing field excitations (particles) which are formally created and absorbed in the asymptotic
vacuum $\ket{0}$ by the external currents $J_i$.
Transition amplitudes, as well as the $S$-matrix which maps initial into final asymptotic states,
do not depend on the explicit parameterisation of the fields $\phi$, but the correlations functions
do depend on the choice of the currents $J_i$, which in turn must be such that $J_i\,\phi^i$
is a scalar under both field redefinitions and changes of spacetime coordinates.
\par
In cosmology, or in the study of the gravitational collapse, the question one would like to answer
is instead what final states $\ket{\zeta_{\rm out},t_{\rm out}}$ are more likely to develop
from an initial state $\ket{\zeta_{\rm in},t_{\rm in}}$ of interest if the field dynamics is driven
by a specific action $S=S[\phi]$ (and then, of course, which states and dynamics
best fit the experimental data).
For example, $\ket{\zeta_{\rm in},t_{\rm in}}$ could be the quantum state representing 
the asymptotically flat regular space of a star (including all information about the matter source),
and one is then interested in computing the probability that this system evolves into a singular
configuration, for which some of the observables develop diverging expectation values at some
$t_{\rm out}>t_{\rm in}$.
The choice of relevant observables to compute depends on the physical system, and one could
consider the energy-momentum tensor of matter for a collapsing star or for a cosmological
model.~\footnote{In cosmology, one is usually interested in the development of
singularities going backward in time, that is for $t_{\rm out}<t_{\rm in}$.} 
In any case, observables will be given by functions of the fields and one should therefore compute
these quantities for all possible final states $\ket{\zeta_{\rm out},t_{\rm out}}$
in order to weigh their relative probability of occurring.
We remark that this problem is further complicated by the fact that the foliation $\Sigma_t$
is also part of what one is investigating, particularly so if one is interested in the possible
development of spacetime singularities.
Strictly speaking, the emergence of the latter would in fact determine the topology of spacetime
by removing (sets of) points from the spacetime manifold.
\par
It should be clear from this brief review of the formalism, that the background field method,
in which one restricts the calculation to transition amplitudes between quantum states
corresponding to the evolution of the background field
\beq
\varphi^i
\coloneqq
\frac{\delta W[J]}{\delta J_i}
\ ,
\label{eqmean}
\eeq
is possibly the only workable assumption.
The definition of the background field~\eqref{eqmean} allows for the introduction of the Legendre
transform
\beq
\Gamma[\varphi]
=
W[J] - J_i \,\varphi^i
\ ,
\label{qS}
\eeq
which satisfies
\beq
\frac{\delta\Gamma[\varphi]}{\delta\varphi^i}
=
-J_i
\ .
\label{eqQeom}
\eeq
When the external current vanishes $J_i=0$, Eq.~\eqref{eqQeom} plays the role of the quantum
generalisation of the Euler-Lagrange equations, which justifies calling $\Gamma[\varphi]$ the effective
(or quantum) action and Eq.~\eqref{eqQeom} its corresponding effective equations of motion.
From Eqs.~\eqref{Fp} and \eqref{qS}, one finds the integro-differential equation for $\Gamma[\varphi]$,
\beq
\exp\left\{i\,\Gamma[\varphi]\right\}
=
\int
\left[\prod_i\d\phi^i\right]
\exp\left\{
\frac{i}{\hbar}
\left[
S[\varphi]
-
\left(\phi^i - \varphi^i\right)
\frac{\delta\Gamma[\varphi]}{\delta\varphi^i}
\right]
\right\}
\ ,
\label{taF}
\eeq
whose exact solution is known only in very simple cases.
In practical terms, one assumes that both the initial state $\ket{\zeta_{\rm in},t_{\rm in}}$
and the final state $\ket{\zeta_{\rm out},t_{\rm out}}$ are such that the corresponding expectation
values of the fields are well approximated by small perturbations $\phi$ about a background
configuration $\varphi$,~\footnote{The background configuration need not be a solution of the
classical equations of motion.
In fact, the background field solves the effective equations of motion in the Schwinder-Keldysh
formalism, in which quantum corrections are taken into account.
The relevance of the background field method therefore goes beyond the study of scattering
amplitudes.}
that is
\be
\phi^i
\to
\varphi^i
+
\phi^i
\ .
\label{bkg}
\ee
The functional machinery can then be used to estimate the expectation values of the
observables by taking functional derivatives of $\Gamma[\varphi]$ with respect to the $\varphi$.
The effective action $\Gamma[\varphi]$ turns out to play a central role in quantum field theory.
For one, it is the generator of one-particle irreducible diagrams, which makes the study of
renormalisation much easier and allows one to readily calculate scattering amplitudes.
Secondly, as we have seen, it generalises the classical action by generating effective
equations of motion for the evolution of the background fields which account
for the backreaction of quantum fluctuations to arbitrary loop order.
\par
For our subsequent analysis, it is important to remark that the dependence
on the quantum states $\ket{\zeta_{\rm in},t_{\rm in}}$ and $\ket{\zeta_{\rm out},t_{\rm out}}$
of the effective action in Eq.~\eqref{taF} is now hidden in the very definition of the background
field, namely
\beq
\varphi^i
=
\bra{\zeta_{\rm out},t_{\rm out}} \phi^i \ket{\zeta_{\rm in},t_{\rm in}}
\ .
\eeq
Like the full quantum generator $W[J]$ would be obtained by functionally integrating over
all field configurations $\phi$ respecting the boundary values implied by $\ket{\zeta_{\rm in},t_{\rm in}}$
and $\ket{\zeta_{\rm out},t_{\rm out}}$, the effective quantum action $\Gamma[\varphi]$ only requires
integrating over the quantum fields (ideally vanishing on both $\Sigma_{t_{\rm in}}$ and $\Sigma_{t_{\rm out}}$).
In either case, the result should not depend on the explicit variables $\phi$ or $\varphi$ we choose
for representing the fields, as long as they also remain spacetime tensors.
The $S$-matrices are indeed invariant under a change of fields.
Nonetheless, the term $(\phi^i-\varphi^i)$ in Eq.~\eqref{taF} is not, thus making the effective action
a non-covariant object under field redefinitions.
\section{The geometry of field space}
\label{Sgeo}
\setcounter{equation}{0}
In order to study objects invariant under field redefinitions, we need to set out the geometry
of field space $\mathcal M$~\cite{DeWitt:1967yk,Isham:1975ur,Giulini:1993ui,Giulini:2009np,
Giulini:1994dx,Giulini:1993ct}, which follows in analogy with Riemannian geometry,
where fields play the role of mere coordinates in the geometrical space $\mathcal M$
(see Ref.~\cite{Parker:2009uva} for a detailed review on the geometrical aspects of the field space).
There is, however, a crucial difference with the usual theory of manifolds regarding the
dimensionality of $\mathcal M$.
Because coordinates in this scenario consist of a set of fields which are
themselves functions of spacetime, $\phi^i = \phi^I(x)$, the field space $\mathcal M$ is infinite-dimensional.
For every fixed spacetime point $x_0\in\Omega$, the space comprised by all $\phi^I(x_0)$ forms
nonetheless a finite-dimensional manifold $\mathcal N$.~\footnote{An example is the space
of non-linear $\sigma$-models.}
As suggested by the notation $\phi^i = \phi^I(x)$, one can thus imagine that the topology
of $\mathcal M$ is given by infinite copies of $\mathcal N$,\beq
\mathcal M = \prod_{x\in\Omega} \mathcal N(x)
\ .
\label{toprod}
\eeq
The above construction concerns only the topological structure of the field space.
Nothing so far has been said about geometrical structures, such as the metric,
nor has it been required.
There is, however, one reason to introduce a metric in field space.
Loop corrections invariably require the calculation of functional determinants of the Hessian
\beq
\det \mathcal H_{ij},
\label{detH}
\eeq
where
\beq
\mathcal H_{ij}
=
\frac{\delta^2 S[\varphi]}{\delta\varphi^i \,\delta\varphi^j}
\ ,
\eeq
denotes the Hessian matrix, which is a bilinear form, {\em i.e.}~it carries two covariant field-space indices.
The determinant of bilinear forms transforms as a tensor density, leading to a dependence on the basis
of the tangent space of $\mathcal M$ in Eq.~\eqref{detH}.
Because the discrete indices in $\mathcal H_{ij}$ generally contain spacetime indices,
not only does the effective action fail to be invariant under field redefinitions, but it
also fails to be invariant under spacetime coordinate transformations.
Determinants of linear operators, {\em i.e.}~objects containing mixed indices, on the other hand,
are invariant under basis transformations.
Thus, to make the determinant of the Hessian invariant under coordinate transformations,
one must transform one of its covariant indices into a contravariant index.
This requires the introduction of a metric in field space.
\par
The metric in $\mathcal M$, hereby denoted $G_{ij}$, must be seen as part of the definition
of the theory, along with the classical action.
The line element is defined as usual as
\beq
\d\mathfrak{s}^2
=
G_{ij}\,\d\phi^i\, \d\phi^j
=
\int_\Omega\d^nx
\int_\Omega\d^nx'
\,G_{IJ}(x,x')\,\d\phi^I(x)\,\d\phi^J(x')
\ .
\eeq
We shall require that $G_{ij}$ be invariant under the same gauge symmetries
used to define the classical action.
This is particularly important to enforce these symmetries at the quantum level
via the path integral measure, which takes a factor $\sqrt{\det G_{ij}}$ to cancel
out the Jacobian determinant from the field redefinition, thus preventing gauge
anomalies.
Apart from symmetry, there is no other guiding principle to help us choose
among all infinite possibilities for a field-space metric.
Nonetheless, should we extend the topological construction~\eqref{toprod}
to geometrical structures, it is natural to assume {\em ultralocality\/}~\footnote{We are
making a slight abuse of notation as $G_{ij} = G_{IJ}(x,x') = G_{IJ}(\phi(x)) \, \delta(x,x')$.
Thus, obviously, $G_{IJ}(x,x')$ and $G_{IJ}(\phi)=G_{IJ}(\phi(x))$ are different objects.
We shall nevertheless use the same tensorial notation but with different arguments
to distinguish them and avoid heavy notations.}
\beq
G_{ij} = G_{IJ} \,\delta(x,x')
\ ,
\eeq
where $G_{IJ}$ depends only on the fields $\phi^I$ and none of their derivatives.
This is an enormous simplification as we reduced the determination of the metric
in the infinite-dimensional field space $\mathcal M$ to the determination of the metric
defined on the finite-dimensional space $\mathcal N$.
Unfortunately, with this level of generality, there are still infinitely many choices
one can make, thus we shall restrict to the simplest cases where one need not include
additional dimensionful parameters.
This means that the field-space metric $G_{IJ}$ will be completely
determined by its tensorial structure, namely the discrete indices contained in $I$ and $J$.
Writing down the most general combination of tensors allowed by the symmetries of $G_{IJ}$,
without however introducing dimensionful coefficients, shall lead to the expression for the
field-space metric.
\par
For a general field space, there is no unique way of defining a
connection over $\mathcal M$ and there is {\em a priori} no reason for adopting
the Levi-Civita connection.
General connections require, however, additional mathematical structures
separate from the metric.
Thus, for simplicity, we shall make again the minimal choice which entails the torsionless
and metric-preserving Levi-Civita connection
\beq
\Gamma^i_{\ jk}
=
\frac12 \,G^{il}
\left(\partial_j G_{kl} + \partial_k G_{jl} - \partial_l G_{jk}
\right)
\ .
\eeq
Our choices imply that the entire geometry of $\mathcal M$ is determined by the field-space
metric $G_{ij}$.
In particular, the ultralocality of the field-space metric extends to the connection
\beq
\Gamma^i_{\ jk}
=
\Gamma^I_{\ JK} \, \delta(x_I,x_J) \,\delta(x_I,x_K)
\ ,
\eeq
where $x_I$ denotes the argument $x$ of the field $\phi^I=\phi^I(x)$ and
\beq
\Gamma^I_{\ JK}
=
\frac12\, G^{IL} 
\left(
\frac{\partial G_{LK}}{\partial\phi^J} 
+
\frac{\partial G_{JL}}{\partial\phi^K} 
- 
\frac{\partial G_{JK}}{\partial\phi^L} 
\right)
\ .
\eeq
The functional Riemannian tensor is defined in the usual way
\beq
\mathcal R^i_{\ jkl}
=
\partial_k \Gamma^i_{\ lj} 
- \partial_l \Gamma^i_{\ kj} 
+ \Gamma^i_{\ km}\,\Gamma^m_{\ lj} 
+ \Gamma^i_{\ lm}\,\Gamma^m_{\ kj}
\ ,
\eeq
with $\mathcal R_{jl} = \mathcal R^i_{\ jil}$ and $\mathcal R = \mathcal R^i_{\ i}$ being
the the functional Ricci tensor and functional Ricci scalar, respectively.
Let us note that, because of the assumption of ultralocality, many contractions
will diverge as $\delta(x,x)$.
This is rather expected and only reflects the infinite dimension of field space
\beq
{\rm dim}(\mathcal M)
\equiv
G_{ij}\, G^{ij}
=
N \, \mathcal V_{(\Omega)} \, \delta(x,x)
\ ,
\eeq
where $N = {\rm dim}(\mathcal N)$ is the dimension of the finite-dimensional space
$\mathcal N$ and $\mathcal V_{(\Omega)}$ denotes the infinite volume of the spacetime
$\Omega$.~\footnote{One could define the effective action on domains
$\bar \Omega\subseteq\Omega$ of finite volume but we shall not go into these details here.}
To make sense of contracted quantities, one must then define densities,~\footnote{One could
also formally deal with undefined products of the Dirac delta by regarding it
as the limit of a sequence of functions.
One can then perform all calculations before taking the limit to the Dirac distribution.}
such as
\beq
\frac{\mathcal R_{ijkl}\, \mathcal R^{ijkl}}{{\rm dim}(\mathcal M)}
=
\frac{1}{N \, \mathcal V_{(\Omega)}} 
\int_\Omega\d^n x\, R_{IJKL}\, R^{IJKL}
\ ,
\eeq
in which divergences due to the infinite dimension of field space, stemming from
both the $\delta(x,x)$ and the infinite spacetime volume, are canceled out.
This clearly has nothing to do with singularities in field space as it is a property of
every theory and not of a specific field configuration.
Since the aforementioned densities can always be defined, we shall implicitly employ
the above procedure and focus on contracted quantities of the finite-dimensional space
$\mathcal N$.
In the following, we shall exemplify the above formalism with non-abelian Yang-Mills
theories and gravitational theories.
\par
For $SU(N_g)$ gauge theories in flat spacetime, the fields are represented by
$\phi^A(x) = A^{a\mu}(x)$ and our minimal choice for the metric leads to
\beq
G_{ij}
=
\eta_{\mu\nu}\, \delta_{ab}\, \delta(x,x')
\ ,
\label{YM}
\eeq
where $a,b = 1, 2, \ldots, N_g^2 - 1$ and $\eta_{\mu\nu}$ is the Minkowski metric.
Note that the only tensors available are $\eta_{\mu\nu}$, $\delta_{ab}$
and $A^{a\mu}$.
However, any other combination of them would require the introduction of a dimensionful
coefficient in order to keep the correct dimensions of $G_{ij}$.
For example, the combination $\alpha \,A_{a\mu}\, A_{b\nu}$ requires a dimensionful parameter
$\alpha$.
The minimal choice has led to a field-independent metric $G_{ij}$ with a trivial geometry,
{\em i.e.}~vanishing connection and curvature.
The metric~\eqref{YM} is incidentally also obtained from the kinetic term of the classical action.
Vilkovisky has indeed suggested that $G_{IJ}$, and ultimately $G_{ij}$, be identified
from the highest-order minimal operator present in the classical action~\cite{Vilkovisky:1984st}.
\par
For metric theories of gravity, one identifies $\phi^I(x) = g^{\mu\nu}(x)$.~\footnote{A common
convention in the literature is to take $\phi^I(x) = g_{\mu\nu}(x)$ with spacetime covariant rather
than contravariant indices.
This generally leads to confusion as field-space covariant indices correspond to spacetime
contravariant indices and vice-versa.}
The assumption of simplicity, together with the symmetries of $G_{IJ}$,
then lead to the one-parameter family of field-space metrics
\beq
G_{ij}
=
\frac{1}{2}
\left(
g_{\mu\rho}\, g_{\sigma\nu}
+ g_{\mu\sigma}\, g_{\rho\nu}
+ c\, g_{\mu\nu}\, g_{\rho\sigma}
\right)
\delta(x,x')
\ ,
\label{eq:dw}
\eeq
which involves only a dimensionless parameter $c$.
The coefficients of the first two terms in Eq.~\eqref{eq:dw} are determined by
requiring $G_{IJ}$ to be a spacetime tensor that satisfies the invertibility condition $G_{IJ} \,G^{JK} = \delta^K_I$.
Such a metric first appeared in the literature in DeWitt's seminal paper~\cite{DeWitt:1967ub}
on the canonical quantisation of gravity.
Its inverse is found by solving $G_{ij} \,G^{jk} = \delta^k_i$, which gives
\beq
G^{ij}
=
\frac{1}{2}
\left( 
g^{\mu\rho}\,g^{\sigma\nu} 
+ g^{\mu\sigma} \,g^{\rho\nu} 
-
\frac{2\, c}{2 + n \, c} \, g^{\mu\nu} \,g^{\rho\sigma}
\right)
\delta(x,x')
\ ,
\eeq
where $n$ is the spacetime dimension.
Note that the DeWitt metric is only invertible for $c\neq -2/n$.
The parameter $c$ cannot be determined without some additional assumption.
The aforementioned procedure proposed by Vilkovisky~\cite{Vilkoviskii:1984un}
gives $c=-1$ for the Einstein-Hilbert action, but it would be different for higher-derivative gravity.
We stress, once again, that Vilkovisky's procedure is not a necessary requirement,
thus we shall leave $c$ unspecified.
For the DeWitt metric, the Ricci tensor is given by
\beq
\mathcal R_{IJ}
=
\frac14
\left(
g_{\mu\nu}\,g_{\rho\sigma} - n \, g_{\mu(\rho}\,g_{\sigma)\nu}
\right)
\label{eq:dwricci}
\eeq
and the Ricci scalar reads
\beq
G^{IJ}\, \mathcal R_{IJ}
=
\frac{n}{4} - \frac{n^2}{8} - \frac{n^3}{8}
\ .
\label{eq:dwriccis}
\eeq
\par
It is standard practice in General Relativity to analyse curvature invariants, like the Kretschmann scalar,
in order to decide whether a singularity is physical or just a coordinate singularity.
Since diffeomorphism invariants are the same in all coordinate systems, only ``true'' singularities
would affect them.~\footnote{Singularities in the scalars derived from the Riemann tensor also
signal the possible divergence of tidal forces, which makes them particularly relevant for physics.}
Analogously, we can seek a scalar functional in order to investigate the appearance of singularities
in the field space.
We could, for example, consider the functional Kretschmann scalar,  
\beq
\mathcal K
=
\mathcal R_{IJKL}\,\mathcal R^{IJKL}
\ ,
\eeq
to assess whether a singularity is real or only a consequence of a bad choice of field variables.
In Ref.~\cite{Casadio:2020zmn}, we have indeed calculated $\mathcal K$ for the DeWitt metric and
found
\beq
\mathcal K
=
\frac{n}{8}
\left(
\frac{n^3}{4} + \frac{3\,n^2}{4} - 1
\right)
\ .
\label{EqK}
\eeq
The fact that $\mathcal K$ is finite everywhere suggests the absence of covariant singularities
in theories of gravity without matter.
At this point, we should stress again that the metric~\eqref{eq:dw} is not unique.
In fact, any metric of the form
\beq
G_{ij}
=
\frac{1}{2}\,(-g)^\epsilon 
\left( g_{\mu\rho}\, g_{\sigma\nu}
+ g_{\mu\sigma}\, g_{\rho\nu}
+ c \, g_{\mu\nu}\, g_{\rho\sigma}
\right)
\delta(x,x') 
\ ,
\label{dwgen}
\eeq
with $g = \det g_{\mu\nu}$, would satisfy ultralocality and simplicity for any value of $\epsilon$.
The choice $\epsilon = 0$ leads to the functional measure originally proposed by Misner~\cite{Misner:1957wq},
whereas the case with $\epsilon=1/2$ was put forward by
DeWitt~\cite{DeWitt:1962ud,DeWitt:1967yk,DeWitt:1965jb}.
Nevertheless, any value of $\epsilon$ is {\em a priori} allowed at the classical level~\cite{Hamber:2009zz}.
The functional Kretschmann scalar for an arbitrary $\epsilon$ turns out to be given by
\beq
\mathcal K
=
\frac{n}{8}\, (-g)^{-2\epsilon}
\left(
\frac{n^3}{4} + \frac{3\,n^2}{4} - 1
\right)
\ .
\label{eq:general}
\eeq
Notice that a covariant singularity in $\mathcal K$ would be present where $g = 0$,
for $\epsilon > 0$, or where $|g|\to\infty$, for $\epsilon < 0$.
Therefore, the case $\epsilon=0$ is the only possibility which excludes a singularity in $\mathcal K$,
regardless of the spacetime metric.
Since the field-space metric is part of the definition of the theory, one could in principle
take the case $\epsilon = 0$ as the definition of a gravitational theory at the classical level.
Nonetheless, at the quantum level, the path integral measure $\sqrt{\det G_{ij}}$ is expected
to preserve the diffeomorphism symmetry of gravity~\cite{Fujikawa:1979ay,Fujikawa:1984qk,
Fujikawa:1983im,Fujikawa:1980vr,Mazur:1990qc,Bern:1990bh,Mottola:1995sj},
thus requiring $\epsilon=1/2$.
In any case, it is not clear whether a singularity in $\mathcal K$ at, say, $g_{\mu\nu}=g^{\rm s}_{\mu\nu}$
would be a real issue.
Removing $g^{\rm s}_{\mu\nu}$ from the field-space manifold $\mathcal M$ and defining covariant
singularities in terms of geodesic incompleteness could, in fact, only mean that $g^{\rm s}_{\mu\nu}$
cannot be realised in Nature.
Contrary to the usual picture of spacetime singularities, which face the philosophical impasse
of the sudden termination of physics, covariant singularities would only signify the absence of
a certain field configuration.
\par
A more useful definition of singularities should be given in terms of physical observables.
Since all observables of a quantum field theory can be computed directly from the quantum action,
$\Gamma[\varphi]$ is a natural candidate to seek a proper and useful definition for singularity.
One then needs a covariant formulation of the effective action, which will be reviewed in the next
section.
Although, in principle, there can be solutions with non-singular observables but with a singular
functional Kretschmann scalar $\mathcal K$, they would not have any practical consequence
for physics and their dangerousness would rather be an epistemological question.
We shall discuss the differences between these types of singularities in depth in Section~\ref{Ssingularity}.
\section{The Vilkovisky-DeWitt effective action}
\setcounter{equation}{0}
\label{Saction}
High-energy physicists are very rarely interested in quantities other than $S$-matrix
elements and cross-sections.
The $S$-matrix contains in fact all accessible information of scattering processes
performed by colliders.
These objects, being defined on-shell, are invariant under field redefinitions that
interpolate between a fixed choice of asymptotic states~\cite{Rebhan:1987cd}.
The importance of scattering amplitudes in cosmology is however secondary.
Unfortunately, one does not have the same level of control to throw cosmological
particles against each other and observe the output.
Observational cosmology is largely based on the measurement of statistical correlation
functions and one is thus rather interested in the evolution of off-shell quantities 
resulting from the backreaction of quantum fluctuations.
Quantum effects for both on-shell and off-shell quantities are however contained
in the same single object: the effective action.
Although $S$-matrices are guaranteed to be invariant under field redefinitions,
off-shell correlation functions are not, which would imply the possible existence
of a preferred field parameterisation.
\par
As we remarked in the Introduction, this is not a problem at the tree level because
the classical action is manifestly a scalar under the field redefinitions~\eqref{eq:1-1}.
Nonetheless, the one-loop effective action acquires a new term
proportional to the classical equations of motion~\cite{Vilkovisky:1984st},
\beq
\Gamma[\varphi]
\to
\Gamma'[\varphi]
=
S[\varphi]
+
\frac{i\,\hbar}{2}
\Tr
\log\left[
\frac{\delta^2 S}{\delta\varphi^m\,\delta\varphi^n}
+
\frac{\delta f^i}{\delta\varphi^m}\,
\frac{\delta f^k}{\delta\varphi^n}\,
\frac{\delta^2 \varphi^l}{\delta f^i\, \delta f^k}\,
\frac{\delta S}{\delta\varphi^l}
\right]
+
\mathcal O(\hbar^2)
\ ,
\label{eq:acttrans}
\eeq
in which we employed the notation of Eq.~\eqref{bkg} and denoted the background
fields with $\varphi$.
The additional term of order $\hbar$ is not important for scattering amplitudes, since
${\delta S}/{\delta\varphi}=0$ on-shell.
However, it does become relevant for the application to cosmology where one
is interested in the off-shell evolution of the mean field.
It is also natural to expect that the invariance of the classical action extends into
the quantum regime.
There is in fact no reason {\em a priori} to prefer one field parameterisation over
the others.
This will be particularly important in the study of singularities under field redefinitions,
as we shall see in the next section.
\par
One way to overcome this issue is to use the geometrical apparatus of Section~\ref{Sgeo}
in order to enforce the invariance of the effective action under field redefinitions.
This was indeed the approach adopted by Vilkovisky~\cite{Vilkovisky:1984st}, where
a connection in the field space is introduced to compensate for the second term
between brackets in Eq.~\eqref{eq:acttrans}.
With a connection in field space, one should then replace the functional derivative
with the corresponding covariant functional derivative
\beq
\frac{\delta^2 S}{\delta\varphi^m\delta\varphi^n}
\to
\nabla_m\nabla_n S
\coloneqq
\frac{\delta^2 S}{\delta\varphi^m\,\delta\varphi^n}
-
\Gamma^i_{\ mn}\,\frac{\delta S}{\delta\varphi^i}
\ ,
\label{eq:subs}
\eeq
and modify the definition of the effective action accordingly.
Since Vilkovisky's effective action only works for a flat field space, DeWitt later on 
proposed the improved effective action~\cite{DeWitt:1988dq}
\beq
\exp\left\{\frac{i}{\hbar}\,\Gamma[\varphi]\right\}
=
\int\d\mu[\phi]\,
\exp\left\{
\frac{i}{\hbar}
\left(
S[\phi]
-\sigma^i(\varphi,\phi)\,(C^{-1})^j_{\ i}[\varphi]\,\nabla_{j}\Gamma[\varphi]
\right)
\right\}
\ ,
\label{eq:qadef}
\eeq
where
\beq
\sigma^i(\varphi,\phi)
=
\frac12
\left(\int_{\gamma(\varphi,\phi)}\d\mathfrak{s}\right)^2
\eeq
is the geodetic interval (which is analogous to Synge's world function~\cite{J.L.Synge:1960zz}),
calculated along the geodesic $\gamma$ with end-points
$\varphi$ and $\phi$, and $C^i_{\ j} = \langle\nabla_j\sigma^i(\varphi,\phi)\rangle_T$.
The angular brackets here denote the functional average, which, for any functional
$F[\varphi,\phi]$, is given by
\beq
\langle F[\varphi,\phi]\rangle_T
=
\exp\left\{-\frac{i}{\hbar} \,\Gamma[\varphi]\right\}
\int\d\mu[\phi]\,
F[\varphi,\phi]\,
\exp\left\{
\frac{i}{\hbar}
\left(S[\phi] + T^i[\varphi,\phi]\, \nabla_i\Gamma[\varphi]
\right)
\right\}
\ ,
\eeq
where $T^i[\varphi,\phi] = \sigma^i(\varphi,\, \phi)(C^{-1})^j_{\ i}[\varphi]$.
Note that the definition of $C^i_{\ j}$ is recursive because $C^i_{\ j}$ also appears
in the definition of the functional average.
Finding an explicit expression for $C^i_{\ j}$ is thus utterly difficult and generally
requires expansions in series ({\em e.g.}~the loop expansion).
The geodetic interval $\sigma^i(\varphi,\phi)$ transforms as a vector at $\varphi$
and as a scalar at $\phi$, thus making the effective action invariant under both
redefinitions of the background $\varphi$ and of the quantum field $\phi$.
The functional measure for a gauge theory reads \cite{Parker:2009uva}
\beq
\d\mu[\phi]
=
[\d\phi]
\,
\mathcal M[\phi;\chi]
\label{eq:measure}
\eeq
with
\beq
[\d\phi]
:=
\prod_k\d\phi^k
\left|\det G_{ij}(\phi)\right|^{1/2}
\eeq
and
\beq
\mathcal M[\phi;\chi]
:=
\left(\det Q^\alpha_{\ \beta}\right)
\tilde\delta[\chi^\alpha]
\ ,
\eeq
where $\det Q^\alpha_{\ \beta}$ is the Faddeev-Popov determinant with
$Q^\alpha_{\ \beta} = \chi^\alpha_{\ ,i}[\phi]\, K^i_\beta[\phi]$ defined in terms
of the gauge fixing $\chi^\alpha$ and the generator of gauge transformations $K^i_\beta$.
The functional Dirac distribution $\tilde\delta[\chi^\alpha]$ is defined analogously to the
standard case as
\beq
\int\left(\prod_\alpha \d\chi^\alpha\right)
\tilde\delta[\chi^\alpha]
=
1.
\eeq
As usual, the gauge fixing ensures that the domain of integration in the functional integral
be restricted to the orbit space, namely the space of distinct equivalence classes unrelated
by a gauge transformation.
The gauge fixing is thus chosen so to pick up only one member of each equivalence class,
which is tantamount to demanding that
\beq
\chi^\alpha[\phi_\epsilon] = \chi^\alpha[\phi],
\label{Gfix}
\eeq
where $\phi^i_\epsilon = K^i_\alpha\, \delta\epsilon^\alpha$,
has only the solution $\delta\epsilon^\alpha = 0$ for a given $\phi^i$.
From this requirement, by expanding the left-hand side of Eq.~\eqref{Gfix} to first order
in $\delta\epsilon^\alpha$, it is straightforward to show that $Q^\alpha_{\ \beta}$ must satisfy
\beq
\det Q^\alpha_{\ \beta} \neq 0.
\label{detQ}
\eeq
The presence of the Faddeev-Popov determinant $\det Q^\alpha_{\ \beta}$
and the gauge condition $\chi^\gamma=0$ imposed by the functional Dirac distribution
guarantee the gauge invariance of the measure.
The determinant of the field space metric in the measure is crucial for obtaining
a path integral measure invariant under field reparameterization.
The covariant measure together with $\sigma^i(\varphi,\phi)$ and the covariant
functional derivative then make the quantum action $\Gamma[\varphi]$ a scalar
functional under redefinitions of both the background $\varphi^i$ and the quantum
field $\phi^i$.
In the one-loop approximation, Eq.~\eqref{eq:qadef} leads to
\beq
\Gamma[\varphi]
=
S[\varphi]
+
\frac{i\,\hbar}{2}\,
\Tr
\log \nabla^i\nabla_j S[\varphi]
+
\mathcal O(\hbar^2)
\eeq
as expected.
We see that the replacement \eqref{eq:subs} has been automatically accounted for.
\par
It is important to emphasize the distinction between the configuration-space
measure (see Eq.~\eqref{eq:measure} above) and the phase-space measure that is used in the
Hamiltonian path integral.
This is indeed a subtle issue that has been the subject of disagreement in the
literature~\cite{Unz:1985wq,Toms:1986sh,Moretti:1997qn,Hatsuda:1989qy,
vanNieuwenhuizen:1989dx,Armendariz-Picon:2014xda,Becker:2020mjl,Buchbinder:1987vp,Hamamoto:2000ab}.
In Ref.~\cite{Unz:1985wq}, Unz had argued that a residual contribution shows up in
the configuration-space measure after integrating out the generalized momenta.
Toms, however, rebutted Unz's argument in Ref.~\cite{Toms:1986sh} by showing that
the configuration-space measure actually does not receive any further contribution
because factors of $g^{00}$ in Unz's result are canceled out by similar factors that
should be included in the phase-space measure from the onset. 
It should also be noted that, contrary to Unz's result, the configuration-space measure
obtained by Toms is explicitly covariant.
The problem is nonetheless quite complicated as the functional measure is only well-defined
after regularization, thus the lack of a rigorous mathematical formulation makes this issue
difficult to settle.
Furthermore, this discussion is clearly limited to Hamiltonians quadratic in the generalized
momenta, for which the momentum integral is Gaussian and can be performed exactly.
The situation is much more difficult in higher-derivative theories where, to our knowledge,
the equivalence between the Lagrangian and Hamiltonian path integrals remains an open
question~\cite{Buchbinder:1987vp,Hamamoto:2000ab}.
In principle, one should therefore view the Lagrangian and Hamiltonian path integrals as different
formalisms that could potentially lead to different predictions in some elaborate situations.
In this case, one could formally choose the configuration-space metric in order to compensate
for the momentum integration and then define the phase-space measure accordingly.
Unless otherwise stated, we shall adopt the Lagrangian path integral as our starting point.
As we shall see later, the differences with respect to Unz's measure are minimal in quadratic
theories.
\par
The Vilkovisky-DeWitt effective action, as defined in Eq.~\eqref{eq:qadef}, is not
automatically the generator of one-particle irreducible (1PI) diagrams.
An additional improvement was made in Ref.~\cite{Ellicott:1987ir}, where it was
shown that in order for the Vilkovisky-DeWitt effective action to generate
1PI diagrams, it requires the generalised definition of the background field
\beq
v^i[\varphi]
=
\frac{\delta W[J]}{\delta J_i}
\ ,
\eeq
thus implying the modified Legendre transform
\beq
\Gamma[\varphi]
=
W[J]
-J_i\, v^i[\varphi]
\ .
\eeq
Such a modification is indeed expected as the current $J_i$ in Eq.~\eqref{qS}
couples to the factor $(\phi^i-\varphi^i)$ in Eq.~\eqref{taF}.
The effective equations of motion then becomes~\footnote{Since $\Gamma$
is a scalar under field redefinitions, we have that
$\nabla_i\Gamma=\partial_i\Gamma\equiv\delta\Gamma/\delta\varphi^i$.}
\beq
\nabla_i\Gamma[\varphi]
=
-J_k\, \nabla_i v^k[\varphi]
\ .
\label{Geom}
\eeq
This modification will play a crucial role in the topological study of
Section~\ref{Stop}.
\par
The quest for singularities now amounts to calculating the on-shell quantum action
$\Gamma[\varphi]$ at any desirable solution of the full effective equations of motion.
Whilst this approach is very general, finding the exact effective equations of motion
(or even the effective action itself) is rather non-trivial.
In practice, one employs approximative methods to obtain the effective action and
its corresponding equations of motion and solutions.
In the semi-classical approximation, for example, one can calculate the solutions order by order
in a loop expansion and singularities can then be studied at each order.
Note that the meaning of going on-shell in this case depends on the loop order
under consideration.
Scattering amplitudes are typically worked out on the mass shell, {\em i.e.}~when
the classical equations of motion are valid.
On the other hand, the on-shell quantum action refers to the quantum action evaluated
on the solution of the quantum equations of motion.
This means that the on-shell quantum action at $n$-loop order is evaluated on the
solution of the quantum equations of motion at $(n-1)$-loop order.
\section{Functional singularities}
\setcounter{equation}{0}
\label{Ssingularity}
With a proper definition for the effective action, we can investigate singularities
in field space whose covariance is now manifest.
We shall define a covariant singularity $\varphi=\varphi_0$ as a solution of the effective
equations of motion in which the Vilkovisky-DeWitt effective action $\Gamma[\varphi_0]$
evaluated at that point is undefined, that is $\Gamma[\varphi_0]$ does not attain
any value or is divergent.
As we have already anticipated, this might happen for two different reasons.
On one hand, the field space $\mathcal M$ might be geodesically incomplete,
in which case $\varphi_0$ would correspond to a point at the boundary of
$\mathcal M$.
Because $\varphi_0$ does not belong to $\mathcal M$, the effective action
$\Gamma[\varphi]$, which takes values from $\mathcal M$ to the real numbers,
is obviously undefined at $\varphi_0$.
This type of covariant singularity thus reflects the absence of certain field configuration
$\varphi_0$, which, as we shall see, is not particularly worrying.
On the other hand, covariant singularities might also be manifested as
a result of the path integral in Eq.~\eqref{eq:qadef}.
In this case, the covariant singularity $\varphi_0$ does belong to $\mathcal M$ but
corresponds to an existing configuration with undefined observables.
In this case, $\varphi_0$ shall be called a functional singularity.
\par
From the definition~\eqref{eq:qadef}, one can see that a functional singularity occurs
whenever the path integral measure~\eqref{eq:measure} either vanishes identically
or diverges at some configuration $\phi=\phi_0$.
From Eq.~\eqref{detQ}, these conditions take place whenever
\beq
[\d\phi](\phi_0)
=
\prod_i \d\phi^i\,\sqrt{|\det G_{ij}[\phi_0]|}
=
\begin{cases}
0
\\
\infty
\ .
\end{cases}
\label{singcond}
\eeq
Since $[\d\phi]$ is the functional measure of a non-gauge theory, the same
conditions~\eqref{singcond} apply regardless of the presence of gauge symmetry.
Under the assumption that the Jacobian of any field redefinition is regular
({\em i.e.}~finite and non-zero), the former possibility translates into
\beq
\sqrt{|\det G_{ij}[\phi]|}
=
0
\ ,
\quad
\forall \phi\in\mathcal M
\ .
\label{eqdet0}
\eeq
Because this condition must be valid for all field configurations, the singularity
would be a property of the entire field-space geometry rather than of some
pathological configuration.
This case is easily remedied by a proper choice of $G_{ij}$ which satisfies
$\sqrt{|\det G_{ij}[\bar\phi]|} \neq 0$ for at least one field configuration $\phi=\bar\phi$.
Therefore, the condition~\eqref{eqdet0} is not a useful proxy for investigating singularities.
\par
On the other hand, the divergence of the measure at a single field configuration
$\phi=\phi_0$ (again for regular Jacobians)
\beq
\lim_{\phi\to\phi_0}\sqrt{|\det G_{ij}[\phi]|} 
=
\infty
\label{fsing}
\eeq
is a sufficient condition~\footnote{Note that the complex exponential in the integrand
of the path integral is bounded.} for the singular behaviour of observables.
From Eq.~\eqref{eq:measure}, the presence of a functional singularity implies that
path integrals and their corresponding observables, such as correlation functions
and $S$-matrix elements, are undefined.
As opposed to the standard singularities in spacetime, whose existence can depend
on the chosen field variables and might not affect the observables after all, functional
singularities not only make observables undefined, but also make the entire path integral
formalism meaningless.
Needless to say, this type of singularities is far more dangerous than the typical
ones in spacetime. 
\par
Note that our definition of functional singularity is not based on the geodesic completeness
of field space.
It is rather a direct way of formalising under what conditions observables are well-defined.
The geodesic completeness has nonetheless important consequences for the formalism.
In fact, if the field space $\mathcal M$ is geodesically incomplete, then the geodetic interval
$\sigma(\varphi,\phi)$ used in the definition~\eqref{eq:qadef} of the Vilkovisky-DeWitt
effective action does not exist for all points in $\mathcal M$ and consequently the
Vilkovisky-DeWitt effective action is not defined everywhere in the field space.
The question of whether a finite-dimensional manifold is geodesically complete
depends on the signature of the metric defined upon it.
For Euclidean metrics, geodesic completeness is guaranteed by the Hopf-Rinow
theorem which, however, does not hold in infinite dimensions~\cite{Atkin}.
On the other hand, for pseudo-Riemannian metrics we have the
Hawking-Penrose theorem~\cite{Hawking:1973uf}, but whether this result can be extended
to the infinite-dimensional field space is an open question (see Appendix~\ref{focus}
for a derivation of the focusing theorem in field space).
Overall, geodesic completeness is not guaranteed in field space and the best we can do
is to search for singularities in the curvature invariants case by case,
as we did for the DeWitt metric (see Eqs.~\eqref{EqK} and \eqref{eq:general}),
in addition to looking for boundaries of the field-space metric $G_{ij}$.
Let us recall that $G_{ij}$ is part of the definition of the theory,
thus one can play this game until a geodesic-complete metric is found.
\par
Despite the technical (as opposed to physical) issue in the definition of the
Vilkovisky-DeWitt effective action, geodesic incompleteness leads to no serious
issues for physics.
For one, contrary to geodesic incompleteness in spacetime, where particles
would cease to exist, there is no physical motion associated to geodesics in field space.
One could for example interpret a point where a geodesic line terminates in field space
as a configuration that cannot be realised in nature.
Such a configuration simply does not contribute to the path integral.
This interpretation is tantamount to a principle of locality in field space:
only nearby configurations in the vicinity of a certain configuration $\phi$ would have
measurable effects upon $\phi$.
Moreover, a straightforward remedy for the effective action on geodesically incomplete
field spaces can be achieved by replacing the geodesic in the definition
of $\sigma(\varphi,\phi)$ by some other curve such that its tangent vector
at $\varphi^i$ is proportional to $(\varphi^i - \phi^i)$.~\footnote{This is required
in order to recover the standard effective action in the limit of flat field space.}
Finally, we should stress that all observables do remain well-defined and finite
everywhere in geodesically incomplete field spaces.
\par
One natural question is whether functional singularities are at all related to
spacetime singularities.
The explicit calculation of the determinant of $G_{ij}$ indeed reveals the possible
relation between functional and spacetime singularities.
In practice, this will largely depend on the choice for field space metric $G_{ij}$.
In the following, we shall give examples of the simplest ({\em i.e.}~with no new
dimensionful parameter) choices for each type of field, namely scalar fields,
abelian and non-abelian gauge fields, spinors and the metric field of gravity.
We shall impose the symmetries present at the classical level on the field-space
metric as well.
Because of gravity, this will demand the presence of the spacetime
volume element $\sqrt{-g}$ for every field.
In particular, this will lead to the choice $\epsilon=1/2$ in Eq.~\eqref{dwgen}.
\par
The simplest choice for a scalar field theory, with $\phi^i = \phi(x)$, would then be
\beq
G_{ij}^{\rm s}
=
\sqrt{-g} \, \delta(x,x')
\ ,
\label{eqsc}
\eeq
where $g = \det g_{\mu\nu}$ is the determinant of the spacetime metric.
Any other choice for the field space metric would invariably introduce 
additional dimensionful parameters.
Calculating the determinant of $G_{ij}^{\rm s}$ explicitly from Eq.~\eqref{eqsc}
points at the relation between singularities in the spacetime $\Omega$
and functional singularities
\beq
\det G_{ij}^{\rm s}
=
\prod_{x\in\Omega} \sqrt{-g}
\ .
\label{detS}
\eeq
Functional singularities thus correspond to divergences of $g$,
which in turn implies that at least one of the eigenvalues of the spacetime
metric is singular.
It would thus appear that spacetime singularities with $g\to\infty$ would not be
removable by field redefinitions.
Nonetheless, because the spacetime metric in this case is not dynamical
(we are just quantising a scalar field in curved spacetime), the path integral
measure is constant in field space.
Quantities of interest are ratios of path integrals, thus the divergences
$g\to\infty$ are canceled out by the normalisation factor.
Such a cancelation will always take place when gravity is treated as external,
as we shall now see for the other types of matter fields.
\par
For a Dirac spinor $\phi^i = (\psi^\alpha(x), \bar\psi^\alpha(x))$, the most
obvious choice for a metric would be~\footnote{Here Greek letters denote spinor indices.}
\beq
G_{ij}^{\rm f} = \sqrt{-g}
\left(
\begin{matrix}
0 & \varepsilon_{\alpha\beta} \\
\varepsilon_{\alpha\beta} & 0
\end{matrix}
\right)
\delta(x,x')
\ ,
\eeq
where $\varepsilon_{\alpha\beta}$ is the inverse of the two-dimensional Levi-Civita tensor.
Its determinant then reads
\beq
\det G_{ij}^{\rm f}
=
\prod_{x\in\Omega}
(-g)^D
\ ,
\eeq
where $D$ is the spinor dimension.
\par
For an abelian Yang-Mills theory, the field is $\phi^i = A^\mu(x)$ and the simplest
field-space metric reads
\beq
G_{ij}^{\rm YM}
=
\sqrt{-g}\, g_{\mu\nu}\, \delta(x,x')
\ ,
\eeq
whose determinant is
\beq
\det G_{ij}^{\rm YM}
=
\prod_{x\in\Omega}
(-1)^{\frac{n}{2}}\, g^{\frac{n}{2} - 1}
\ .
\eeq
For a Yang-Mills theory with gauge group $SU(N_g)$, one identifies the field-space
coordinates as $\phi^A(x) = A^{a\mu}(x)$.
We adopt the field space metric
\beq
G_{ij}^{\rm nYM}
=
\sqrt{-g}\, g_{\mu\nu}\, \delta_{ab}\, \delta(x,x')
\ ,
\label{eq:ym}
\eeq
which is the simplest choice in this case.
The determinant can then be readily calculated to be
\beq
\det G_{ij}^{\rm nYM}
=
\prod_{x\in\Omega}
(-1)^{\frac{n}{2}(N_g^2-1)}\, g^{\frac12 (n-2)(N_g^2-1)}
\ .
\eeq
\par
Note that none of the above field-space metrics depend on the dynamical fields
({\em i.e.}~$\partial_k G_{ij} = 0$), thus the Levi-Civita connection and consequently
the Riemann tensor vanish.
Moreover, their determinants are constant in field space, resulting in a factor that
gets trivially canceled out by the normalisation factor in the path integral.
As we have anticipated, spacetime singularities do not propagate to the observables
which are themselves finite even for singular spacetime metrics.
It is astonishing that this happens even at the classical level.
This picture might however change when gravity is quantised, which is the subject
of the following.
\par
In the case of gravitational theories, the determinant of the DeWitt metric~\eqref{dwgen}
(with $\epsilon=1/2$) reads
\beq
\det G_{ij}^{\rm DW}
=
\prod_{x\in\Omega}
(-1)^{n-1} \left(1 + \frac{c\, n}{2}\right)g^{\frac14 (n-4)(n+1)}
\ ,
\eeq
and again we can see the relation between functional singularities and spacetime
singularities for any spacetime dimension $n$.
Amusingly, in four dimensions $\det G_{ij}$ becomes constant for any metric configuration,
suggesting the absence of singularities in this case.
Apart from the four-dimensional case, functional singularities do appear for singular spacetime
metrics.
Notice that the spacetime metric is no longer external, thus one can no longer cancel it out by
the normalisation factor of path integrals.
We should also stress that the DeWitt metric accounts only for the gravitational sector.
In real systems, matter is nonetheless always present, which is expected to lead to different
conclusions.
A complete field-space metric, corresponding to a theory with $N_{\rm s}$ scalars, $N_{\rm f}$ fermions,
$N_{\rm YM}$ abelian gauge fields, $N_{\rm nYM}$ non-abelian gauge fields and the metric,
can be constructed by laying out the above metrics along its diagonal,
\beq
G_{ij}
=
{\rm diag}\,
\big(
\overbrace{G_{IJ}^{\rm s}, \ldots, G_{IJ}^{\rm s}}^{N_{\rm s}},
\overbrace{G_{IJ}^{\rm f}, \ldots, G_{IJ}^{\rm f}}^{N_{\rm f}},
\overbrace{G_{IJ}^{\rm YM},\ldots, G_{IJ}^{\rm YM}}^{N_{\rm YM}},
\overbrace{G_{IJ}^{\rm nYM},\ldots, G_{IJ}^{\rm nYM}}^{N_{\rm nYM}},
G_{IJ}^{\rm DW}
\big) \, \delta(x,x')
\ .
\label{fullmetric}
\eeq
The resulting determinant then reads~\footnote{We recall, as noted before, the difference
between the configuration-space and phase-space measures.
In view of Toms' result~\cite{Toms:1986sh}, Eq.~\eqref{detGfinal} does not receive any further
modifications from the momentum integration.
On the other hand, following Unz~\cite{Unz:1985wq}, Eq.~\eqref{detGfinal} would receive an additional
multiplicative factor of $(g^{00})^\zeta$.
The power $\zeta$ depends on the number of degrees of freedom,
being positive for bosons and negative for fermions.
The presence of functional singularities in this case would then have to be inferred from the behaviour
of $(g^{00})^\zeta\, (-g)^\beta$ instead.}
\beq
\det G_{ij}
=
\prod_{x\in\Omega}
(-1)^\alpha
\left(1+\frac{cn}{2}\right)
(-g)^\beta
\ ,
\label{detGfinal}
\eeq
where
\begin{align}
\alpha
&=
\frac{1}{2}\,N_{\rm s}
+
D\,N_{\rm f}
+
\frac{n}{2}\,N_{\rm YM}
+
\frac{n}{2}
\left(N_g^2 - 1\right)
N_{\rm nYM}
+
n - 1
\\
\beta
&=
\frac{1}{2}\,N_{\rm s}
+
D\,N_{\rm f}
+
\frac{(n - 2)}{2}\,N_{\rm YM}
+
\frac{(n - 2)}{2} \left(N_g^2 - 1\right)
N_{\rm nYM}
+
\frac{1}{4} (n - 4)(n + 1)
\ .
\end{align}
Therefore, the presence of functional singularities in an arbitrary theory
is parameterised by $\beta$, which depends solely on the particle content
of the model.
Note, in particular, that $\beta$ is strictly positive for $n \geq 4$,
thus functional singularities could be present in a theory of quantum gravity coupled to matter
with an arbitrary action and whose field-space metric is given by Eq.~\eqref{fullmetric}.
This conclusion is nonetheless dependent on the actual choice of field-space metric.
Avoiding singularities thus require deviating from the simplest diagonal choice made in
Eq.~\eqref{fullmetric}.
We should stress that the condition~\eqref{fsing} imposed on the geometry of field-space
is a sufficient but not necessary condition for the presence of functional singularities.
In fact, the final outcome of the path integral depends on global contributions and boundary
conditions in addition to the local value of the functional measure.
This makes Eq.~\eqref{fsing} a good proxy to reveal functional singularities,
but not a good one to infer their absence.
For that, we need topological methods that we study in the next section.
\section{Topological classification of functional singularities}
\label{Stop}
\setcounter{equation}{0}
In the last section, we have laid out the connection between functional singularities
and the geometry of field space.
We shall now relate functional singularities to the topology of maps between the field space
and the real circle $\mathbb S^1$, as suggested by the definition of the effective action.
This will lead to the classification of functional singularities in terms of a winding number
defined in the field space $\mathcal M$, which allows for the elaboration of some strategies
to obtain regular quantum field theories.
For the purposes of this section, we set $\hbar = 1$ for simplicity.
\par
From Eq.~\eqref{eq:qadef}, it is natural to define the functional
\beq
\psi[\varphi]
\coloneqq
e^{i\,\Gamma[\varphi]}
\ ,
\label{orderp}
\eeq
in order to investigate functional singularities.
Indeed, the presence of a functional singularity at $\varphi=\varphi_0$
translates into $\psi[\varphi_0] = 0$ or $\psi[\varphi_0] = \infty$, at which points the
effective action becomes undefined.
In fact, Eq.~\eqref{orderp} has no solutions for $\Gamma[\varphi_0]$
for these values of $\psi[\varphi_0]$, thus the effective action at $\varphi_0$ does not exist.
We shall call $\psi[\varphi]$ the functional order parameter because $\psi$ plays
the analogous role of an order parameter in the theory of phase transitions in ordered media
or cosmology~\cite{Mermin:1979zz,Hindmarsh:1994re,Kibble:1999yk}.
The field space $\mathcal M$ can be thought of as the ordered medium itself,
whereas functional singularities correspond to topological defects.
Let us recall, however, that $\mathcal M$ is an infinite-dimensional space.
Assuming that $\Gamma$ is real,~\footnote{There exist cases where the effective action is complex,
which usually signals an instability.
Such cases are however trivial and do not contain functional singularities.
In fact, the functional order parameter space for a pure imaginary effective action is the set of
non-negative real numbers $\mathbb{R}_{\geq 0}$, whereas for a generally complex effective
action is the disk $D^2 \subset \mathbb{C}$.
Both these spaces are however simply connected $\pi_1(\mathbb{R}_{\geq 0}) = \pi_1(D^2) = 0$.}
the functional order parameter $\psi$ defines the map
\beq
\psi : \mathcal M \to \mathbb S^1,
\eeq
from the field space to the unit circle, the latter playing the role of the
order parameter space.
If we encircle an exact solution $\varphi_0$
with a $d$-dimensional hypersurface $\gamma_d(\varphi_0) \subset \mathcal M$ whose topology
is $\mathbb S^d$,
the functional order parameter restricted to $\gamma_d(\varphi_0)$ induces the map
\beq
\psi|_{\gamma_d} : \mathbb S^d \to \mathbb S^1
\eeq
between higher-dimensional spheres centred at $\varphi_0$ and the circle.
Covariant singularities can thus be classified using the homotopy groups
$\pi_d(\mathbb S^1)$.
\par
Apart from the fundamental group $\pi_1(\mathbb S^1) = \mathbb{Z}$,
which is isomorphic to the integers, all higher homotopy groups of the circle are trivial,
namely $\pi_d(\mathbb S^1) = 0$ for $d>1$.
This means that all hyperspheres in $\mathcal M$ can be continuously contracted
to a point with the exception of the circle $\gamma_1$.
The latter may find obstructions that prevent it from being continuously
contracted to a point.
Such obstructions are precisely the functional singularities defined before.
They could be given by strings (or higher-dimensional submanifolds)
in field space, {\em i.e.}~extended higher-dimensional objects,
along which the effective action is undefined.
Since $\pi_1(\mathbb S^1) = \mathbb Z$, the homotopy classes are labeled by the
winding number~\footnote{This definition of the winding number holds for a functional singularity
with $\psi = 0$.
Nonetheless, the winding number around $\psi \to \infty$ can be analogously obtained
by shifting the order parameter $\psi[{\varphi}] + \psi_0 = e^{i\Gamma[\varphi]}$
with $\psi_0 \to \infty$.
The final result turns out to be the same Eq.~\eqref{wind2}.}
\begin{align}
\mathcal{W}
&=
\frac{1}{2\,\pi\,i}
\oint_{\psi[\gamma_1]} 
\frac{\delta\psi}{\psi}
\nonumber
\\
&=
\frac{1}{2\,\pi}
\int_0^{2\,\pi}
\d\theta
\int_\Omega\d^n x\, 
\left.\frac{\partial \mathcal L(x)}{\partial \varphi^I(x)}\right|_{\varphi^I(x)=\gamma^I(x;\theta)}
\frac{\d\gamma^I(x;\theta)}{\d\theta}
\ ,
\label{wind2}
\end{align}
where $\psi[\gamma_1]$ denotes the image of $\gamma_1$ under the map
$\psi[\varphi]$.
The field configurations $\gamma^i=\gamma^I(x;\theta)$ are an explicit parameterisation
of $\gamma_1$ in terms of the angle $0\le\theta\le 2\,\pi$ such that
$\gamma^I(x;0)=\gamma^I(x;2\,\pi)$ and, of course,
\be
\Gamma[\varphi]
=
\int_\Omega
\d ^n x\,\mathcal{L}(\varphi^I,\partial_\mu\varphi^I,\ldots)
\ ,
\ee
with $\mathcal{L}$ the effective Lagrangian density.
Note that, because $\delta\Gamma=\delta\psi/\psi$ is an exact form, the winding number does not depend
on the curve $\gamma_1$.
\par
A functional singularity exists whenever $\mathcal W \neq 0$.
Note that, in general, the closed curve $\gamma_1$ runs over points $\varphi \in \mathcal M$
that are not solutions of the exact equations of motion.
On the other hand, if one can find a closed curve $\gamma_1$ about the point $\varphi_0$
that is restricted only to solutions of the full equations of motion~\eqref{Geom} with $J_i=0$,
then Eq.~\eqref{wind2} implies that $\varphi_0$ is not singular.
Similarly, if there is a curve $\gamma_1$ whose tangent vector is everywhere normal~\footnote{Of course,
this explicitly depends on the field-space metric $G_{ij}$.}
to $\delta\Gamma/\delta \varphi^i$, then $\varphi_0$ is non-singular.
In principle, Eq.~\eqref{wind2} establishes a well-defined procedure to determine whether
a certain field configuration is singular.
\par
Computing the effective action in general, however, is far from trivial and one usually needs
to rely on approximate methods.
We shall now see a simple example where we can compute $\mathcal W$ and then move
to the case of gravity.
In these examples, we do not calculate the effective action from first principles,
we rather assume that it has been given.
This is, however, enough to show the power of our formalism. In this sense, both examples
should be considered as toy models.
Whether these toy models are realised in real systems does not concern us here.
\subsection{Scalar fields in four dimensions}
\label{SecEx1}
Let us consider a field theory with two real scalars $\varphi^i = (\varphi^1(x), \varphi^2(x))$
and the effective action
\beq
\Gamma[\varphi^1, \varphi^2]
=
\int_\Omega\d^4x
\left[\frac12\, \partial_\mu\varphi^1\, \partial^\mu\varphi^1
+
\frac12\, \partial_\mu\varphi^2\, \partial^\mu\varphi^2
-
V(\varphi^1,\varphi^2)\right]
\ ,
\label{acS}
\eeq
where the potential
\beq
V(\varphi^1,\varphi^2)
=
\Lambda^4\,
\arctan\!\left(
\frac{\varphi^1}{\varphi^2}
\right)
\ ,
\label{EqPot}
\eeq
and $\Lambda$ is a mass parameter accounting for the correct dimensions.
Note that the potential, thus the effective action, is undefined at the trivial solution
$\varphi^I_0(x) = (0,0)$, which makes such a point a field singularity by our definition.
It is then convenient to consider $\gamma_1$ given by homogeneous and static configurations 
encircling the origin $\varphi^i_0 = (0, 0)$, that is
\beq
\gamma^I(x;\theta)
=
\left(A\, \cos\theta, A\, \sin\theta\right)
\ ,
\label{circ}
\eeq
where $A$ is an arbitrary positive constant.
The variation of the effective action yields
\beq
\frac{\delta\Gamma[\varphi^1, \varphi^2]}{\delta\varphi^I(x)}
=
-\Box\varphi^I(x) - \frac{\partial V}{\partial\varphi^I(x)}
\ ,
\eeq
which, when evaluated along $\gamma_1$, gives
\begin{align}
\frac{\delta\Gamma}{\d\theta}
&=
\int_\Omega
\d^4 x
\left.
\frac{\partial\mathcal L}{\partial\varphi^I}\right|_{\varphi^I=\gamma^I}
\frac{\d\gamma^I}{\d\theta}
\nonumber
\\
&=
\int_\Omega
\d^4 x
\left(
\left.\frac{\partial V(\varphi)}{\partial \varphi^1(x)}\right|_{\varphi^I(x)=\gamma^I}
\!\!\!\sin\theta
-
\left.
\frac{\partial V(\varphi)}{\partial \varphi^2(x)}\right|_{\varphi^I(x)=\gamma^I}
\!\!\!\cos\theta
\right)
A
\nonumber
\\
&=
\mathcal{V}_{(4)}\,\Lambda^4
\ ,
\label{var}
\end{align}
where $\delta\Gamma/\d\theta$ denotes the functional derivative along the curve $\gamma_1$
and $\mathcal{V}_{(4)}$ is the 4-volume of the whole spacetime $\Omega$.
Note that the kinetic terms vanish along $\gamma_1$, because we are considering
static and homogeneous configurations (which, furthermore, are not solutions of the effective
equations of motion).
Note also that, for this simple case, the same result can be obtained from the direct calculation
of the effective action on the encircling configurations~\eqref{circ}, that is
\be
\Gamma(\theta)
=
-\mathcal{V}_{(4)}\,\Lambda^4\,\arctan(\cot\theta)
=
\mathcal{V}_{(4)}\,\Lambda^4\,\theta
\ ,
\ee
and then use
\be
\mathcal W=
\frac{1}{2\,\pi}
\int_0^{2\,\pi}
\frac{\delta\Gamma}{\d\theta}\,
\d\theta
\ .
\label{wind3}
\ee
Finally, the winding number \eqref{wind2} is simply given by
\beq
\mathcal W = 1
\ ,
\label{WS}
\eeq
provided we set $\mathcal{V}_{(4)}=\Lambda^{-4}$.
Let us recall that the winding number is an integer by definition, thus we must choose
$\mathcal{V}_{(4)}$ and $\Lambda^{-4}$ to comply with this fact.
This is formally reflected in the normalisation of the parameter $\theta$
along $\gamma_1$ as an angle.
We should note that, because different theories have different couplings,
eliminating the IR divergence due to an infinite volume would require different choices
for the relation between $\mathcal  V_{(4)}$ and the coupling constants in the theory
in order to achieve this normalisation.
For our purposes, what matters is that the formalism for functional singularities can distinguish
between field spaces with $\mathcal W = 0$ everywhere and those with $\mathcal W \neq 0$
for paths encircling specific configurations, non-zero winding numbers with different magnitudes
being fundamentally equivalent.
\par
The above calculation shows that the field configuration $\varphi^i_0 = (0,0)$ is indeed a functional
singularity of non-zero topological charge.
Note that in obtaining this result we assumed that the effective action had already
been calculated and handed to us in the closed form \eqref{acS}.
The calculation of the effective action, however, depends on the geometry of field
space via the path integral measure.
It is clear from Eq.~\eqref{detS} that the functional singularity \eqref{WS} could not
have resulted from the divergent measure for the simplest field-space metric~\eqref{eqsc},
because $\det G^{\rm s}_{ij}$ is regular in flat spacetime.
It is also important to stress that a non-zero winding number is a necessary and
sufficient condition for the existence of functional singularities, whereas a divergent
$\det G_{ij}$ is only sufficient.
Therefore, the result \eqref{WS} could either reflect a more complicated field-space
metric than \eqref{eqsc}, whose determinant diverges, or a result that is not
captured solely by the path integral measure.
\subsection{Scalar field in FLRW spacetime}
\label{SecEx2}
A typical example of a spacetime singularity in which the determinant of the spacetime
metric vanishes, namely $g=0$, is the system of a homogeneous massless scalar field $\phi=\phi(t)$
minimally coupled to the Friedmann-Lemaitre-Robertson-Walker (FLRW) metric.
For simplicity, we shall assume the effective action
\beq
\tilde \Gamma 
=
\int_\Omega
\d^4x \, \sqrt{-g}
\left(
	\frac{R}{16\, \pi \,\gn}
	- \frac12\, \partial_\mu \phi \,\partial^\mu \phi
\right),
\label{eaS}
\eeq
where $\gn$ denotes Newton's constant and $R$ is the Ricci scalar.
In the ADM decomposition, the spatially flat FLRW metric is given by
\beq
\d s^2
=
-N^2 \,\d t^2
+ a^2 \left[(\d x^1)^2 + (\d x^2)^2 + (\d x^3)^2\right]
\,
\label{FLRW}
\eeq
where we have set the shift functions $N_i$ to zero,
$N=N(t)$ denotes the lapse function and $a=a(t)$ the scale factor.
The Ricci scalar for the metric \eqref{FLRW} is given by
\beq
R
=
\frac{6}{N^2}
\left(
	\frac{\ddot{a}}{a}
	- \frac{\dot{a} \,\dot{N}}{a\, N}
	+ \frac{\dot{a}^2}{a^2}
\right)
\,
\label{Rs}
\eeq
where we adopted the standard notation $\dot{a} = \d a/\d t$ for time derivatives.
Plugging the FLRW metric~\eqref{FLRW}, along with~\eqref{Rs}, in the effective action~\eqref{eaS}
leads to
\begin{align}
\Gamma
\equiv
\frac{\tilde\Gamma}{\mathcal{V}_{(3)}}
&=
\frac{1}{2}
\int
\d t
\left[
\frac{1}{\kappa}
\left(
\frac{a^2\, \ddot a}{N}
-
\frac{a^2\,\dot a\,\dot N}{N^2}
+
\frac{a\,\dot a^2}{N}
\right)
+
\frac{a^3\,\dot\phi^2}{N}
-
\frac{\d F}{\d t}
\right]
\nonumber
\\
&=
-\frac12
\int\d t
\left(
 \frac{a \, \dot{a}^2}{\kappa\, N}
- \frac{a^3 \,\dot{\phi}^2}{N}\right)
\equiv
\int\d t\, L
\ ,
\label{gammaFLRW}
\end{align}
where we have defined $\kappa=8\,\pi\,\gn/6$ and $\mathcal{V}_{(3)}$ is the spatial volume
corresponding to the spatial isometries of the chosen field configurations, in analogy with the
previous example.
Note that, in obtaining Eq.~\eqref{gammaFLRW}, we have included the total derivative of
\be
F
=
\frac{a^2\,\dot a}{\kappa\,N}
\label{total}
\ee
in order to remove second derivatives of $a$, which further eliminates $\dot N$.~\footnote{In fact, $N$
is not a dynamical variable but a Lagrange multiplier corresponding to the time-reparameterisation
invariance of the model.}
The variation of the Lagrangian then reads
\be
\delta L
&\!\!=\!\!&
\frac{a^2}{\kappa\, N}
\left(
	-\frac{\ddot{a}}{a}
	- \frac{\dot{a}^2}{2\,a^2}
	+ \frac{\dot{a}\, \dot{N}}{a\, N}
	- \frac{3}{2}\,\kappa\, \dot{\phi}^2
\right)
\delta a
-
\frac{\d}{\d t}
\left(
	\frac{a^3\, \dot{\phi}}{N}
\right)
\delta\phi
\nonumber
\\
&&
+
\frac12
\left(
	\frac{a\, \dot{a}^2}{\kappa\,N^2}
	- \frac{a^3 \,\dot{\phi}^2}{N^2}
\right)
\delta N
\ ,
\label{varS}
\ee
which gives the equations of motions of each degree of freedom when $\delta L = 0$.
For $N=1$, this system has solutions of the form
\begin{align}
a^3_\pm(t)
&=
\pm 3 \sqrt{\kappa} \, p_\phi \, t
\label{asol}
\\
\phi_\pm(t)
&=
\pm \frac{1}{\sqrt{\kappa}}\,
\log\left(\pm \frac{t}{t_0}\right)
\ ,
\label{psol}
\end{align}
where $t_0$ is an integration constant and $p_\phi = a^3\, \dot{\phi}$ is a constant of motion
that follows from the equation for $\phi$.
Note that we have taken $p_\phi$ positive for simplicity, since its sign does not affect our analysis.
The positive (negative) sign of the scale factor $a$ in Eq.~\eqref{asol} corresponds to an expanding
(contracting) universe with $0 < t < \infty$ (respectively, $-\infty < t < 0$).
For simplicity, we have also chosen the integration constants for $a$ such that $a_\pm(0) = 0$.
This allows us to join the contracting and expanding solutions at $t=0$ to form the ``bouncing''
configuration which we denote as $\varphi^i_{\rm s}=(a_{\rm s}(t),\phi_{\rm s}(t))$ for brevity.
Note that the Ricci curvature~\eqref{Rs} for the solutions~\eqref{asol} diverges for $t\to 0$,
indicating the presence of a spacetime singularity at the bounce in $t = 0$, where furthermore the scalar field
$|\phi_{\rm s}|\to \infty$. 
Notably, the effective action~\eqref{gammaFLRW} is however finite for the solutions~\eqref{asol}-\eqref{psol}
(indeed $\Gamma[\varphi_{\rm s}]=\Gamma[a_\pm,\phi_\pm]=0$), suggesting that the spacetime singularity
at $t = 0$ has no corresponding functional singularity.
In the following, we shall confirm this by showing that the winding number for the bouncing configuration
$\varphi_{\rm s}$ is indeed equal to zero.
\par
Let us recall that to calculate the winding number for a certain field configuration,
we must encircle that specific configuration with a curve $\gamma_1$.
Similarly to the previous example, we shall parameterise $\gamma_1$ around $\varphi_{\rm s}$
as
\beq
\gamma^I(t;\theta)
=
\left(a_{\rm s}(t) + A\,\cos\theta, \phi_{\rm s}(t) + A \,\sin\theta, 1\right)
\ ,
\label{parS}
\eeq
for all values of $t$ for which $a=a_{\rm s}(t)$ and $\phi=\phi_{\rm s}(t)$ are defined, and $A$
is a positive constant.
Note that the Lagrangian $L$ in Eq.~\eqref{gammaFLRW} diverges (like $t^{-2}$) for $t\to 0$
when computed on the above configurations (for $\theta$ fixed).
This makes the calculation of the effective action $\Gamma$ along the encircling configurations~\eqref{parS}
quite tricky because the time integral in Eq.~\eqref{gammaFLRW} diverges.
In order to avoid such complications, we can exploit the freedom to add total derivatives
to $L$~\footnote{Let us recall that
total derivatives affect neither the (effective) equations of motion nor scattering amplitudes.}
and, in particular, we replace $F$ in Eq.~\eqref{total} by
\beq
F
=
\alpha\, \dot \phi + \beta \,\dot a
\eeq
and take
\begin{align}
\alpha &= \frac{A^3}{3}\, \cos^3\theta
\\
\beta &= \frac{A^2}{2}\, \cos^2\theta
\ .
\end{align}
This choice for the total derivative cancels out the divergence in the time integral on the
configurations~\eqref{parS},~\footnote{One could also regularise
the time integral in Eq.~\eqref{gammaFLRW} with a cut-off $|t|>T>0$ and take $T\to 0$ at the end.
It can be easily verified that this procedure also leads to $\mathcal W = 0$.}
while keeping $\Gamma[\varphi_{\rm s}]=0$. 
Moreover, we remark that $F(\gamma^I(t;\theta))\to 0$ for $t\to\pm\infty$.
The resulting effective action evaluated along~\eqref{parS} then vanishes identically, namely
\beq
\Gamma(\theta) = 0
\,
\eeq
and $\mathcal W = 0$ follows.
As we expected, the spacetime singularity at $t=0$ does not correspond to a functional singularity.
The most striking consequence of this result is that all physical observables remain finite and well-defined
for the bouncing solution,~\footnote{We must emphasise that physical effects cannot be
changed by a change of field variables.
In particular, the fact that physical observables remain finite at spacetime singularities does not preclude
strong tidal forces.
The {\em spaghettification\/} indeed takes place regardless of the chosen frame being smooth or not.}
even at $t=0$.
This indeed reflects the existence of field-space coordinates in which the spacetime singularity
completely disappears~\cite{Kamenshchik:2016gcy, Kamenshchik:2017ojc, Kamenshchik:2018crp}.
Note that this conclusion cannot be reached from the functional measure associated
to \eqref{detS}, as the vanishing of the measure at a single configuration is not sufficient
to yield a vanishing path integral. 
The fact that $\mathcal W = 0$ indeed results from non-trivial global contributions to the path integral.
\subsection{Winding number and external currents}
The winding number $\mathcal W$ can be expressed in terms of the external current,
which brings the possibility of evaluating $\mathcal W$ without the need of $\Gamma[\varphi]$.
In fact, using the effective equations of motion~\eqref{Geom} in the expression~\eqref{wind2}
for the winding number, we find
\beq
\mathcal W
=
-\frac{1}{2\,\pi}
\oint_{\gamma_1} 
J_k \, v^k_{\ ;i} \, \d\gamma^i
\ .
\eeq
From the Stokes theorem, one then obtains~\footnote{A rigorous definition of functional integration
is a well-known open problem in mathematics, thus the formal application of Stokes theorem to
infinite-dimensional spaces must be carried out with great care.
In our case, we recall that $\gamma_1$ is finite-dimensional (and so is ${\mathcal A}$), thus one
can apply the Stokes theorem as usual.}
\begin{align}
\mathcal W
&=
-\frac{1}{2\,\pi}
\int_{\mathcal A}
(J_k \, v^k_{\ ;i})_{,j} \, \d\varphi^j\wedge\d\varphi^i
\nonumber
\\
&=
-\frac{1}{2\,\pi}
\int_{\mathcal A}
\xi_{ij} \, \d\varphi^j\wedge\d\varphi^i
\ ,
\label{wind4}
\end{align}
where ${\mathcal A}$ is a surface with boundary $\partial {\mathcal A} = \gamma_1$ and 
we used the antisymmetry of the wedge product, with
\beq
\xi_{ij}[\varphi]
\coloneqq
(J_k \, v^k_{\ ;i})_{,j}
-
(J_k \, v^k_{\ ;j})_{,i}
\ .
\eeq
We can now proceed as before and parameterise the surface ${\mathcal A}$ by $\zeta^I(x; r, \theta)$
in terms of the radius $r > 0$ and the angle $0 \le \theta < 2\,\pi$.
In this parameterization, Eq.~\eqref{wind4} becomes
\beq
\mathcal W
=
-\frac{1}{2\,\pi}
\int_{\mathcal A}
\d r \wedge \d\theta
\int\d^n x
\int\d^n x'
\,
\xi_{IJ}[r,\theta]
\,
\frac{\partial\zeta^I(x; r, \theta)}{\partial r}
\frac{\partial\zeta^J(x'; r, \theta)}{\partial\theta}
\ ,
\label{wind5}
\eeq
with
\begin{align}
\xi_{IJ}[r,\theta]
=
\int\d^n x^K
&
\left\{
\frac{\delta}{\delta \varphi^J(x^J)}\bigg[J_K(x^K) \, \nabla_{\varphi^I(x^I)} v^K(x^K)\bigg]_{\varphi^I(x^I) = \zeta^I(x^I; r, \, \theta)}
\right.
\\
&
\left.
-
\frac{\delta}{\delta \varphi^I(x^I)}\bigg[J_K(x^K) \, \nabla_{\varphi^J(x^J)} v^K(x^K)\bigg]_{\varphi^I(x^I) = \zeta^I(x^I; r, \, \theta)}
\right\}
\ ,
\end{align}
where we denoted $i = (I, x^I)$ to keep track of the different indices.
Recall that both $J_I(x) = J_I[\varphi(x)]$ and $v^I(x) = v^I[\varphi(x)]$ are functionals of $\varphi^i$.
This expression can be further simplified if we assume the standard effective
equations, namely $v^k_{\ ;i}\equiv \nabla_i v^k =\delta^k_{i}$,
in which case we obtain
\beq
\xi_{IJ}[r, \theta]
=
\left.
\frac{\delta J_I(x^I)}{\delta \varphi^J(x^J)}
\right|_{\zeta^I(x^I; r, \, \theta)}
-
\left.
\frac{\delta J_J(x^J)}{\delta \varphi^I(x^I)}
\right|_{\zeta^I(x^I; r, \, \theta)}
\ .
\label{rotJ}
\eeq
This is however only valid when the field-space geometry satisfies $\mathcal R_{ijkl}\,v^l = 0$,
which includes the case of a flat field space.
From Eqs.~\eqref{wind5} and \eqref{rotJ}, we see that $\mathcal W$ can be interpreted as the flux
of the curl of the external current $J_i$ across the area determined by the circuit $\gamma_1$.
Note that the vanishing curl of the external current, $\xi_{IJ}[r,\theta] = 0$,
is a sufficient (but not necessary) condition for the absence of singularities.
\par
The above topological consideration shows that there is an infinite number of possible
functional singularities, each one labeled by the winding number $\mathcal W$.
These are potential singularities, but their actual presence depends on the specifics of the
path integral and field-space geometry, as outlined in Section~\ref{Ssingularity}.
Given the effective action, one is then able to determine explicitly whether its corresponding
theory contains functional singularities as we did in the examples of Sections~\ref{SecEx1}
and \ref{SecEx2}.
The crucial point in such a topological classification is that one can now assess whether
spacetime singularities, predicted by the Hawking-Penrose theorem, lead to any catastrophic
consequence for the theory itself. 
As we showed in Section~\ref{SecEx2}, it is indeed possible that known spacetime singularities
turn out to be removable when $\mathcal W = 0$.
Because the Vilkovisky-DeWitt effective action is invariant under field redefinitions, so is the
winding number.
The absence of functional singularities in combination with the invariance of the winding number
under field redefinitions suggests the existence of some coordinates in field space in which the
spacetime singularity is instead regular.
The formalism developed in this section can thus be used to enforce the vanishing
of the winding number, $\mathcal W=0$, to mitigate functional singularities and, ultimately,
spacetime singularities.
This might indeed be helpful in the construction of a regular field-space metric,
in relation to the description of functional singularities in Section~\ref{Ssingularity}.
\section{Conclusions}
\label{Sconc}
\setcounter{equation}{0}
Spacetime singularities have since long been pointed as one of the reasons General Relativity
needs replacement.
The generality of the Hawking-Penrose theorem makes it difficult to overcome spacetime
singularities even in modified gravity theories, with the exception of some very special models.
With the principle of covariance in field space, it becomes crucial to put singularities under the
microscope and analyse them under this new perspective.
As it is somewhat expected, the Hawking-Penrose theorem does not survive field redefinitions.
The calculation of the field-space Kretschmann scalar in pure gravity has also suggested that
classical observables, defined as scalar functionals in field space, are finite for the special case
$\epsilon=1/2$ of the DeWitt metric~\cite{Casadio:2020zmn}.
\par
In this paper, we have taken another step in understanding the meaning of singularities in physics.
We considered the effective action as the onset for the investigation of singularities since
it encodes the information about all physical observables.
A closer look into its definition has revealed two potential types of singular behaviour.
The first, and less dangerous one, takes the form of points where geodesics become incomplete
for a finite value of the affine parameter.
They are thus analogous to spacetime singularities and are the ones that would have been
revealed by the functional Kretschmann scalar.~\footnote{A singular Kretschmann
scalar implies the presence of a covariant singularity but the opposite is not true.
In fact, a singular Kretschmann scalar is a sufficient but not necessary condition for the presence
of singularities, for there are infinitely many invariants on a given manifold that one should assess
before concluding that no singularity is present.}
At the classical level, geodesic completeness appears to be sufficient for the proper
definition of classical observables.
On the other hand, at the quantum regime there is another source of singularity as the
path integral can become divergent or vanish identically, in which case the effective action
cannot be defined.
It is important to stress that such divergences are not UV divergences of perturbative
quantum field theory. Functional singularities are rather defined at the non-perturbative
level and they are reminiscent of divergent path-integral measures which inherit their
properties from the corresponding geometry of field space.
It is somewhat surprising that both the functional Kretschmann scalar and
the path-integral measure remain regular in four spacetime dimensions for the DeWitt metric.
This suggests that $n=4$ stands at a special place from the perspective of the geometry of
field space.
The presence of matter, however, changes this situation considerably as a functional singularity
always exists when the determinant of the spacetime metric $g\to\infty$, unless gravity is treated
classically or semi-classically.
We must again emphasise that the effective action stems from the interplay between the
field-space geometry, the classical action and the boundary conditions in the path integral.
For this reason, a regular field space alone does not guarantee the absence of functional singularities.
\par
The fundamental group of the functional order parameter space, on the other hand, takes into account
all the three ingredients above.
It provides a topological classification of functional singularities, which are then labeled by the winding number.
Whether such functional singularities are really present in a given theory, however, depends
on the resulting effective action, which ultimately hinges on the particular geometry of the field space,
on the classical action as well as on boundary conditions.
Generally, an effective action with vanishing winding number is free of functional
singularities.
We showed that the winding number indeed vanishes for the class of field theories with
functionally irrotational external sources and whose functional Riemann tensor satisfies
$\mathcal R_{ijkl}\,v^l = 0$.
We thus conclude by remarking that the topological classification of functional singularities, along with the
geometry of field space, serves as an important tool in the construction of a consistent
theory of quantum gravity.
\subsection*{Acknowledgments}
I.K., A.K. ~and R.C.~are partially supported by the INFN grant FLAG.
The work of R.C.~has also been carried out in the framework of activities
of the National Group of Mathematical Physics (GNFM, INdAM) and COST
action Cantata.
A.K.~is partially supported by the Russian Foundation for Basic Research 
grant No  20-02-00411.
\appendix
\section{Focusing theorem in field space}
\label{focus}
\setcounter{equation}{0}
The existence of caustics in $\mathcal M$ can be understood in terms of the
convergence of a family of geodesics with the aid of the Raychaudhuri equation
in field space, which ultimately translates into a condition on the functional Ricci tensor.
The proof follows the same logic of the standard result in spacetime,
which we review in the following for the case of Riemannian metrics.~\footnote{Analogous
results can be found for time-like and null-like geodesics in Lorentzian field spaces.}
Let us assume that the field space $\mathcal M$ can be foliated by hypersurfaces
orthogonal to geodesics so that
\beq
\d\mathfrak{s}^2
=
(\d\phi^{0})^2
+ G_{\bar \imath \bar \jmath}\, \d\phi^{\bar \imath}\,\d\phi^{\bar \jmath}
\ ,
\eeq
where $\phi^0$ is a fiducial but otherwise arbitrary direction~\footnote{One can obviously
choose any of the fields $\phi^{I}$, different choices correspond to different foliations of
$\mathcal M$.} taken to be orthogonal to the others (denoted by barred indices).
The condition for the existence of a focal point, namely a point where the geodesic
congruence converges, is given by
\beq
\det G_{\bar \imath \bar \jmath}
=
0
\ .
\label{fcond}
\eeq
Due to the ultralocality of the field-space metric, the focusing condition~\eqref{fcond}
can be translated into a condition on the metric of the finite-dimensional manifold $\mathcal N$,
\beq
\prod_{x\in\Omega}
\det G_{\bar I \bar J}(x)
=
0
\quad
\iff
\quad
\det G_{\bar I \bar J} = 0
\ ,
\eeq
so that a focal point in $\mathcal N$ is also a focal point in $\mathcal M$.
We can thus study the convergence of a geodesic congruence in terms of $G_{\bar I \bar J}$.
Since $G_{\bar I \bar J}$ is the metric of the finite-dimensional space $\mathcal N$,
one can repeat the reasoning used to demonstrate the focusing theorem in spacetime
as we shall now review.
\par
Let $X^{\bar I}$ be the normalized vector field tangent to the geodesic congruence,
parameterised by an affine parameter $\lambda$ along geodesics, and orthogonal
to the hypersurface $\phi=\phi^0$ for a given $\lambda$.
The functional covariant derivative of $X^{\bar I}$ can be split into the irreducible
representations of the group $SO(N)$ as
\beq
\nabla_{\bar I} X_{\bar J}
=
\varsigma_{\bar I \bar J} + \Omega_{\bar I \bar J} + \frac{\vartheta}{N-1}\,h_{\bar I \bar J} 
\ ,  
\eeq
where $N$ is the dimension of $\mathcal N$, 
$h_{\bar I \bar J} = G_{\bar I \bar J} - X_{\bar I} \,X_{\bar J}$ and
\begin{align}
\vartheta
&=
G^{\bar I \bar J}\,\nabla_{\bar I}\, X_{\bar J}
\\
\varsigma_{\bar I \bar J}
&=
\nabla_{(\bar I} X_{\bar J)}
- \frac{\vartheta}{N-1}\,h_{\bar I \bar J} 
\\
\Omega_{\bar I \bar J}
&=
\nabla_{[\bar I} X_{\bar J]}
\end{align}
denote the functional expansion parameter, the functional shear tensor and the functional
twist tensor of the congruence, respectively.
The functional Raychaudhuri equation then reads
\beq
\frac{\delta\vartheta}{\d\lambda}
=
-
\frac{\vartheta^2}{N-1}
-
\varsigma_{\bar I \bar J}\,\varsigma^{\bar I \bar J}
+
\Omega_{\bar I \bar J} \,\Omega^{\bar I \bar J} 
-
\mathcal R_{\bar I \bar J}\,X^{\bar I}\, X^{\bar J}
\ ,
\eeq
where ${\delta}/{\d\lambda}$ is the functional derivative along the congruence.
Assuming an irrotational congruence and noting that $\vartheta^2$ and
$\varsigma_{\bar I \bar J}\,\varsigma^{\bar I \bar J}$ are non-negative, one finds that
\beq
\mathcal R_{\bar I \bar J} \,X^{\bar I} \,X^{\bar J}
\ge
0\label{eq:cond}
\eeq
implies
\beq
\frac{\delta\vartheta}{\d\lambda}
\leq
-\frac{\vartheta^2}{N-1}
\ .
\label{tobeint}
\eeq
Eq.~\eqref{tobeint} can be formally integrated to give
\beq
\vartheta
\leq
-\left(\frac{1}{C} - \frac{\lambda}{N-1}\right)^{-1}
\ ,
\eeq
where $C$ is a constant of integration.
The expansion parameter can be written in terms of $\mathcal V = \sqrt{\det{G_{\bar I \bar J}}}$ as
\beq
\vartheta
=
\frac{1}{\mathcal V}\frac{\delta\mathcal V}{\d\lambda}
\ ,
\eeq
in which case the above inequality reads
\beq
0 \leq \mathcal V \leq \mathcal V(0)\left(1-\frac{C\lambda}{N-1}\right)^{N-1}
\ ,
\eeq
after integration.
This implies that $\mathcal V$ vanishes for $\lambda=(N-1)/C$,
showing that all geodesics starting off orthogonally from the hypersurface
at $\phi=\phi^0$ will eventually converge to a point.
Differently from spacetime, the configuration manifold is not dynamical,
thus $\mathcal R_{\bar I \bar J}$ has to be evaluated case by case to infer
what theories would have focal points.
Note that the above result is not restricted to gravity, so that any field theory
satisfying Eq.~\eqref{eq:cond} necessarily have focal points.
%
%
%
%
%

\end{document}